\documentstyle[aps,prl,epsf,floats,multicol,amssymb,tighten]{revtex}
\begin{document}

\newcommand{\base}{}
\newcommand{\nonb}{\nonumber}
\newcommand{\RA}{\rangle}
\newcommand{\LA}{\langle}
\newcommand{\LL}{\langle \langle}
\newcommand{\RR}{\rangle \rangle}
\newcommand{\HG}{\hat{G}}
\newcommand{\HK}{\hat{K}}
\newcommand{\Ht}{\hat{t}}
\newcommand{\HA}{\hat{A}}
\newcommand{\HI}{\hat{I}}
\newcommand{\HX}{\hat{X}}
\newcommand{\HY}{\hat{Y}}
\newcommand{\Hg}{\hat{g}}
\newcommand{\Hrho}{\hat{\rho}}
\newcommand{\DA}{{\cal D}^A}
\newcommand{\DR}{{\cal D}^R}
\newcommand{\CG}{{\cal G}}
\newcommand{\D}{{\cal D}}
\newcommand{\W}{{{\cal W}_{n,m}(N_a,N_b)}}
\newcommand{\vk}{{\vec{k}}}
\newcommand{\vx}{{\vec{x}}}

\newenvironment{tab}[1]
{\begin{tabular}{|#1|}\hline}
{\hline\end{tabular}}

\newcommand{\fig}[2]{\epsfxsize=#1\epsfbox{#2}} \reversemarginpar 
\bibliographystyle{prsty}

\title{Transport theory of multiterminal hybrid structures}

\author{R. M\'elin$^{(1)}$\thanks{melin@polycnrs-gre.fr}
and D. Feinberg$^{(2)}$}
\address{Centre de Recherches sur les Tr\`es Basses
Temp\'eratures (CRTBT)\thanks{U.P.R. 5001 du CNRS,
Laboratoire conventionn\'e avec l'Universit\'e Joseph Fourier}\\
{CNRS BP 166X, 38042 Grenoble Cedex, France}}
\address{$^{(2)}$ Laboratoire d'Etudes des Propri\'et\'es
Electroniques des Solides\\
{CNRS BP 166X, 38042 Grenoble Cedex, France}}
\maketitle

\begin{abstract}
\normalsize
We derive a microscopic transport theory 
of multiterminal hybrid structures in which
a superconductor is connected to several
spin-polarized electrodes.
We discuss the non-perturbative
physics of extended contacts, and show that it
can be well represented by averaging out the phase
of the electronic wave function. 
The maximal
conductance of a two-channel contact is
proportional to
$(e^2/h) (a_0/D)^2 \exp{[-D/\xi(\omega^*)]}$,
where $D$ is the distance between the
contacts, $a_0$ the lattice spacing,
$\xi(\omega)$ is the superconducting
coherence length,
and $\omega^*$ is the cross-over frequency
between a perturbative regime ($\omega<
\omega^*$) and
a non perturbative regime ($\omega^* < \omega < \Delta$).
The intercontact Andreev reflection and elastic
cotunneling conductances are not equal
if the electronic phases take a fixed value.
However, these two quantities do coincide
if one can average out the electronic phase.
The equality between the Andreev and cotunneling
conductances is also valid in the
presence of at least one extended
contact in which the phases take
deterministic values.
\base
\end{abstract}

\widetext

%%%%%%%%%%%%%%%%%%%%%%%%%%%%%%%%%%%%%%%%%%%%%%%%%%%%%%%%%%%%%%%%%%

\section{Introduction}

Transport of correlated pairs of electrons in multiterminal
configurations 
has recently focused an important interest.
One possible line of research is
motivated by the possibility of creating entangled pairs of
electrons from a
superconductor~\cite{Martin,Loss,Feinberg,Falci,Melin}.
This may lead to fundamental tests of quantum mechanics 
in solid state, or to new ways of manipulating quantum information. In a 
different context, the interplay between
superconductivity and magnetism 
offers novel
functionalities in multiterminal devices : the various transport 
channels occurring at several neighboring superconducting-ferromagnet 
interfaces depend in a subtle way on spin polarizations and
geometry~\cite{deJong,Soulen,Upadhyay,Buzdin,Apinyan}. 
At the theoretical level, there is a need for a transport theory in
multiterminal hybrid structures involving superconducting
and spin-polarized  elements.
These structures
have so far not been the object of any experiment.
As a consequence, one of the objective of the
present time models is to predict what
should be measured in future experiments.

In ferromagnet~-- superconductor junctions,
it is well established that Andreev reflection
is suppressed by an
increase of spin polarization~\cite{deJong}.
This is because Andreev reflection can take
place only in the channels having both
a spin-up and a spin-down Fermi point.
This theoretical prediction has been 
probed experimentally by two independent
groups~\cite{Soulen,Upadhyay}. 
Spin polarized Andreev reflection~\cite{Soulen}
and related effects~\cite{Melin00}
can even be used to measure the Fermi surface
spin polarization.

We are concerned here
with more sophisticated systems in which
a superconductor is connected to several
electrodes, which can be ferromagnetic
or normal metals. In the case of ferromagnetic
electrodes, it will be crucial to take into
account the existence of a very small
coherence length. We neglect any diffusive
effect in spite of the fact that they lead
to a rich physics~\cite{Giroud,Petrashov,Chandra,Bauer}.
As a consequence, our models should apply to
point contacts having a dimension much smaller
than the diffusive mean free path. The fabrication
of such contacts in multiterminal configurations
may seem difficult in view of the present
day technology, and this is why there are
no available experiments on these systems.
However, there are interesting 
phenomena taking place in
these multiterminal systems.
For instance, Andreev reflection
can become {\sl non local}. Namely a
spin-up electron from a given electrode A
can be Andreev reflected as a hole
in a {\sl different} electrode B.
This effect has been studied theoretically
by Byers and Flatt\'e in Ref.~\cite{Byers}
for normal metals, and in Ref.~\cite{Feinberg}
in the ferromagnetic case which contains the
richest physics. It is in our
opinion crucial to develop the most general
theoretical description of this phenomenon.
In this respect, two approaches
have been developed recently. One is based
on the analysis of the lowest order processes
appearing in perturbation theory~\cite{Falci}.
Another approach is non perturbative, but relies on
effective Green's
functions~\cite{Melin}. It is a very
natural task to work out the
microscopic theory of transport in
ballistic
multiterminal hybrid structures.
Transport theory will be solved exactly
by means of Green's function techniques~\cite{Cuevas}.
We incorporate in our theoretical description
two realistic constraints:
\begin{itemize}
\item[(i)] {\sl Multichannel
effects} which are expected to play a central
role in quantum point contacts involving
ferromagnetic metals. The radius of the
contact can be smaller or larger than
the phase coherence length of the ferromagnetic metal.
\item[(ii)] {\sl The strength of the tunnel amplitude}
is small in low transparency contacts,
and large in high transparency contacts.
Our approach is non perturbative. Therefore,
the tunnel matrix element can take arbitrary
values. As a result, we can derive transport
in the presence of arbitrary bias voltages.
\end{itemize}
An ingredient that is not incorporated at the
present stage in the model is the reduction
of the superconducting gap associated to
the proximity effect.

We will use this non perturbative approach to
address the following physical questions:
\begin{itemize}
\item[(i)] It has been already established
that the multiterminal hybrid system should be
described by a {\sl conductance matrix}.
The matrix elements encode all information
about the current flowing in
a given electrode, in response to
a voltage applied in another electrode.
{\sl How does the crossed conductance behave
when the voltages are close to the 
superconducting gap ? What is the maximal
value of the crossed conductance ?}

\item[(ii)] The superconductor Green's functions
contain not only an information about 
non local processes, but contain also
an information about the phase of electron
propagation. In extended contacts, there are
many phases coming into account. {\sl Can these
phases be considered as random quantities ?}

\item[(iii)] There are two propagators associated
to a superconductor: the ordinary and the
anomalous propagators.
One can easily realize that after phase
averaging, the ordinary propagator is identical to
the anomalous propagator (see section~\ref{sec:form-G}).
As a consequence, in the tunnel approach
and for unpolarized contacts,
the averaged Andreev
reflection conductance is equal to the the
averaged elastic cotunneling conductance~\cite{Falci}.
{\sl Is this identity still valid in the
presence of large interface transparencies ?}
\end{itemize}
%%%%%%%%%%%%%%%%%%%%%%%%%%%%%%%%%%%%%%%%%%%%%%%%%%%%%%%%%%%%%

The article is organized as follows.
Some preliminaries are given in section~\ref{sec:summary}.
The form of the Green's functions is derived
in section~\ref{sec:form-GF}.
The solution of the model with two single-channel
electrodes
is presented in section~\ref{sec:two-channel}.
The general solution with an arbitrary number
of single-channel electrodes 
is presented in section~\ref{sec:general}.
As a particular example, we discuss in
section~\ref{sec:3channel} the physics
of a model with three single-channel
electrodes. Multichannel electrodes
are solved in section~\ref{sec:multi-chan}.

\section{Preliminaries}
\label{sec:summary}

\subsection{The Keldysh method}
We will use Green's functions techniques to solve
transport theory. There is an advanced
($\HG^A$), retarded ($\HG^R$)
and Keldysh ($\HG^{+,-}$) Green's
functions~\cite{Keldysh,Caroli}.
Each of these
Green's functions is a $2 \times 2$ matrix
in Nambu representation. The Dyson equation
for the advanced and retarded Green's functions
takes the form
\begin{equation}
\label{eq:Dy1}
\HG^{R,A} = \Hg^{R,A} + \Hg^{R,A} \otimes \hat{\Sigma}
\otimes \HG^{R,A}
.
\end{equation}
The Dyson equation for the Keldysh component is given by
\begin{equation}
\label{eq:Dy2}
\HG^{+,-} = \left[ \HI + \HG^R \otimes \hat{\Sigma}
\right] \otimes \Hg^{+,-} \otimes
\left[ \HI + \hat{\Sigma} \otimes
\HG^{A} \right]
.
\end{equation}
Eqs.~\ref{eq:Dy1},~\ref{eq:Dy2} are written in a compact
notation in which the convolution involves a summation
over time variables and space labels. $\hat{\Sigma}$
is the self energy, which contains all couplings
present in the tunnel Hamiltonian.
The notation $\Hg$ is used for the Green's functions
of the disconnected system ({\sl i.e.} with $\hat{\Sigma}=0$)
while $\HG$ refers to
the Green's functions of the connected system
({\sl i.e.} with $\hat{\Sigma} \ne 0$).
We will use the notation
\begin{equation}
\label{eq:fg}
\Hg^{A,R}(t,t') = \left( \begin{array}{cc}
g^{A,R}(t,t') & f^{A,R}(t,t') \\
f^{A,R}(t,t') & g^{A,R}(t,t') \end{array} \right)
\end{equation}
for the Nambu representation of the
advanced and retarded Green's functions, with
\begin{eqnarray}
g^A(t,t') &=& -i \theta(t-t') \langle \left\{ c_{i,\uparrow}(t) ,
c_{j,\uparrow}^+(t') \right\} \rangle\\
f^A(t,t') &=& -i \theta(t-t') \langle \left\{ c_{i,\uparrow}(t) ,
c_{j,\downarrow}(t') \right\} \rangle
.
\end{eqnarray}
We will also denote by $\hat{\rho}={1 \over \pi} \mbox{Im}(\hat{g}^A)$
the Nambu representation of the density of states:
\begin{equation}
\label{eq:Nambu-rho}
\hat{\rho} = \left( \begin{array}{cc}
\rho_g & \rho_f \\
\rho_f & \rho_g \end{array} \right)
,
\end{equation}
with $\rho_g = {1 \over \pi} \mbox{Im} (g^A)$
and $\rho_f = {1 \over \pi} \mbox{Im} (f^A)$.
The Nambu representation
of a given tunnel matrix element connecting sites $a$ and $\alpha$
is $t_{a,\alpha} \hat{\sigma}^z$, where $\hat{\sigma}^z$
is one of the Pauli matrices.
We use a notation in which
the ``sites'' of the superconductor are represented
by the Greek symbols $\alpha$, $\beta$ , $\gamma$... 
The sites in the normal metal electrodes
are represented by the Latin symbols 
$a$, $b$, $c$ , ... The explicit form of the
Keldysh Green's function connecting the two
sides of a given interface is, from (\ref{eq:Dy2})
\begin{equation}
\label{eq:Gpm-gene}
\HG^{+,-}_{\alpha_k,a_k} = \sum_{i,j}
\left[ \delta_{k,i} \HI
+ \HG^R_{\alpha_k,a_i}
\Ht_{a_i,\alpha_i} \right]
\Hg^{+,-}_{\alpha_i,\alpha_j}
\Ht_{\alpha_j,a_j}
\HG_{a_j,a_k}^A
 + \sum_{i,j}
\HG^R_{\alpha_k,\alpha_i}
\Ht_{\alpha_i,a_i}
\Hg^{+,-}_{a_i,a_j} 
\left[
\delta_{k,j} \HI +
\Ht_{a_j,\alpha_j}
\HG^A_{\alpha_j,a_k} \right]
.
\end{equation}
The strategy is first to use (\ref{eq:Dy1}) to
calculate the advanced and retarded
Green's functions and next
use (\ref{eq:Dy2})
to calculate the Keldysh Green's function.
The current can be obtained easily from the
Keldysh Green's function~\cite{Caroli}:
\begin{equation}
\label{eq:I-gene}
I_{a_k,\alpha_k} = \frac{e}{h}
\int d \omega \left[ \Ht_{a_k,\alpha_k} \HG^{+,-}_{\alpha_k,a_k}
- \Ht_{\alpha_k,a_k} \HG^{+,-}_{a_k,\alpha_k} \right]
.
\end{equation}
The spin-up (spin-down) current is given by
the 11 (22) matrix element of the Nambu representation.

\subsection{A useful trick on the spectral current}
\label{sec:conj-spectral}

The systems of interest here
are made of a single superconductor
connected to an arbitrary number of external 
normal metal electrodes. It turns out that
there exists some tricks that can be used
to simplify the calculation of the current
in such systems.
One of these tricks is the following.

The terms in the first
summation in Eq.~\ref{eq:Gpm-gene}
contain a prefactor
$n_F(\omega-\mu_S)$ because $\alpha_i$ belongs
to the superconductor. The terms in
the second summation contain a
prefactor $n_F(\omega-\mu_{a_i})$
because $a_i$ belongs to a 
ferromagnetic electrode.
The trick consists in realizing that
the terms containing $n_F(\omega-\mu_S)$
coincide exactly with the terms
containing $n_F(\omega-\mu_{\alpha_i})$.
As a result, the total current can be written
as an integral over energy of the spectral current:
\begin{equation}
\label{eq:conj-I}
\label{eq:courant-gene}
I_k = \sum_i \int d \omega \left[ n_F(\omega-\mu_{\alpha_i})
- n_F(\omega-\mu_S) \right] {\cal I}_{k,i}(\omega)
.
\end{equation}

%%%%%%%%%%%%%%%%%%%%%%%%%%%%%%%%%%%%%%%%%%%%%%%%%%%%%%%%%%%%%

\section{Form of the Green's functions}
\label{sec:form-GF}
\label{sec:3.1.3}
In this section, we present a derivation of
the form of the Green's functions that will be used
throughout
the remainder of the article. This will
give us the opportunity to discuss 
the relevant parameters of the model.

\subsection{Green's functions in the superconductor}

\subsubsection{Effective Green's functions}
\label{sec:single-site}
In some cases,
it will be useful to describe the superconducting
and ferromagnetic reservoirs in terms of effective
Green's functions. It was already
shown by one of us in Ref.~\cite{Melin} that effective
Green's functions can be used to construct
a consistent non perturbative ``toy model''
version of transport theory.
In this approach, the
superconductor is viewed as zero dimensional:
its dimensions are shorter than the
coherence length.
The superconducting effective Green's function
takes the form~\cite{Cuevas}:
\begin{equation}
\label{eq:G-single-gen}
\Hg^{R,A}(\omega) = \frac{ \pi \rho_N}
{\sqrt{ \Delta^2 - (\omega-\mu_S)^2}}
\left[ \begin{array}{cc}
- (\omega-\mu_S) \pm i \eta & \Delta\\
\Delta & - (\omega-\mu_S) \pm i \eta
\end{array} \right]
,
\end{equation}
and we will consider
in the following the limit $\eta \rightarrow 0$.
The Keldysh component is given by
$\Hg^{+,-}(\omega) = 2 i \pi n_F(\omega-\mu_S)
\hat{\rho}(\omega)$, with $\hat{\rho}(\omega)
= { 1 \over \pi} \mbox{Im}(\Hg^A)$ the density 
of states.
The ferromagnetic electrodes are described
in a similar way:
\begin{equation}
\label{eq:single-Ferro}
\Hg^{R,A} = \mp i \pi
\left[ \begin{array}{cc}
\rho_{1,1} & 0 \\
0 & \rho_{2,2}\end{array} \right]
,
\end{equation}
where $\rho_{1,1}$ and $\rho_{2,2}$
are respectively the spin-up and
spin-down densities of states.

\subsubsection{Spectral representation}
To address more realistic models, it is useful
to restore the dependence of the Green's functions
upon space coordinates. The Green's function are
evaluated in terms of a spectral
representation. Let us start
with the pairing Hamiltonian 
\begin{equation}
{\cal H} = \sum_{ \vk,\sigma}
\xi_k c_{\vk,\sigma}^+ c_{\vk,\sigma}
+ \Delta_k^* c_{\vk,\downarrow}^+
c_{\vk,\uparrow}^+
+ \Delta_k c_{\vk,\uparrow}
c_{\vk,\downarrow}
,
\end{equation}
with $k=|\vk|$. We use the notation
$\xi_\vk = \epsilon_\vk - \mu$,
with $\epsilon_\vk = \hbar^2 k^2
/ (2m)$ for the kinetic energy.
After standard manipulations, the spectral
representation is found to be
($i,j$ standing for $\alpha_i,\alpha_j$):
\begin{eqnarray}
\label{eq:spec-supra-1}
\left[g_{i,j}^{A}\right]_{1,1}
(\omega) &=&
\frac{1}{\cal N} \sum_k e^{i\vk.(\vx_i-\vx_j)}
\left[
\frac{ (u_k)^2}{\omega - (\mu_S+E_k)+i\eta} +
\frac{ (v_k)^2}{\omega - (\mu_S-E_k)+i\eta}
\right]\\
\label{eq:spec-supra-2}
\left[g_{i,j}^{A}\right]_{1,2}(\omega) &=&
\frac{1}{\cal N} \sum_k e^{i\vk.(\vx_i-\vx_j)}
u_k v_k
\left[
-\frac{1 }{\omega - (\mu_S+E_k) + i \eta}
+\frac{ 1}{\omega - (\mu_S-E_k) + i \eta}
\right]
,
\end{eqnarray}
where ${\cal N}$ the number of sites in the
superconductor. We used the standard notation
\begin{eqnarray}
E_k &=& \sqrt{\Delta^2 + (\xi_k)^2}\\
(u_k)^2 &=& \frac{1}{2}\left( 1 + \frac{\xi_k}{E_k}
\right)\\
(v_k)^2 &=& \frac{1}{2}\left( 1 - \frac{\xi_k}{E_k}
\right)
\end{eqnarray}
for the quasiparticle energy, and the
electron and hole coherence factors.

\subsubsection{Form of the Green's function}
\label{sec:form-G}
The spectral representation (\ref{eq:spec-supra-1}),
(\ref{eq:spec-supra-2}) is valid in any dimension.
We now restrict our discussion to the case of
a three dimensional superconductor.
We first perform the angular integration and
next use the residue theorem to make
the radial integration. Note that it is crucial
to carry out first the angular integration.
This ensures the existence of well defined
convergence properties when we use the
residue theorem to make the radial integration.
The final result is
\begin{eqnarray}
\label{eq:spectral-3D}
\Hg_{i,j}^{R,A}(\omega) &=& 
\frac{m a_0^3}{\hbar^2}
\frac{1}{2 \pi |\vx_i-\vx_j|}
\exp{ \left( - \frac{ |\vx_i-\vx_j|}
{2 \xi(\omega)} \right)}\\
&&\times
\left\{
\frac{\sin{\varphi}}{\sqrt{\Delta^2 - (\omega-\mu_S)^2}}
\left[ \begin{array}{cc}
-(\omega-\mu_S)\pm i \eta &  \Delta \\
 \Delta & -(\omega-\mu_S)\pm i \eta \end{array}
\right]
- \cos{\varphi}
\left[ \begin{array}{cc}
1 & 0 \\ 0 & 1
\end{array} \right] \right\}
,\nonb
\end{eqnarray}
with $\varphi=k_F |x_i - x_j|$
and $a_0$ is the length of the elementary cell.
The coherence
length appearing in (\ref{eq:spectral-3D}) is
\begin{equation}
\label{eq:xi-om}
\xi(\omega) = \left\{
\begin{array}{cc}
\xi(0) \frac{\Delta}{\sqrt{\Delta^2 - \omega^2}}
& \mbox{ if $\omega < \Delta$}\\
+ \infty
& \mbox{ if $\omega > \Delta$}
.
\end{array}
\right.
\end{equation}
We used the notation
$\xi(0)={\epsilon_F \over k_F \Delta}$ for the zero-frequency
coherence length, with $\epsilon_F$ the Fermi energy.
In the case of two point contacts $a$ and $b$ treated
explicitly in section~\ref{sec:two-channel},
the Green's function $g_{a,b}^{R,A}$ provides
a coherent coupling between charge transport
at the two contacts.
We end-up this section with three remarks.
First, we note that with
$\cos{\varphi}=0$ the
Green's functions are identical to the effective Green's
functions given in
section~\ref{sec:single-site}.
Second, we recover the usual free-fermion
Green's function in the limit $\omega \gg \Delta$:
$$
g_{i,j}^{R}(\omega) = -i
\frac{m a_0^3}{\hbar^2}
\frac{1}{2 \pi |\vx_i-\vx_j|}
e^{i k_F | \vx_i - \vx_j|}
.
$$
Finally, we will discuss in detail the role played by
phase averaging. Using the notation $\LL ... \RR = \int
\frac{d \varphi}{2 \pi}$, one
can show that
$\LL \left( g_{i,j} \right)^2 \RR
= \LL \left( f_{i,j} \right)^2 \RR
$.
This identity implies that in the tunnel limit
the average
Andreev reflection conductance is equal to
the average elastic cotunneling conductance.

\subsection{Green's functions in the ferromagnetic
electrodes}
\label{sec:Green-ferro}
The Green's function in the ferromagnetic electrodes
are diagonal in Nambu space.
The form of the Green's function is taken as
\begin{equation}
\label{eq:spectral-ferro}
g_{i,j,\sigma}^{R}(\omega)= -i \frac{m a_0^3}{\hbar^2}
\frac{1}{2 \pi |\vx_i-\vx_j|}
\exp{\left( i \varphi^{(\sigma)} \right)}
\exp{\left(- \frac{|x_i-x_j|}{2 l_{\phi}^{(\sigma)}} \right)}
,
\end{equation}
where the phase is given by
$\varphi^{(\sigma)}=k_F^{(\sigma)} |x_i-x_j| $
and $l_{\phi}^{(\sigma)}$ is the phase coherence length.
There is a mismatch between the spin-up and spin-down
Fermi wave vectors:
$$
k_F^{(\sigma)} = \frac{\sqrt{2 m}}{\hbar}
\sqrt{\epsilon_F + \sigma h_{\rm ex} + \omega}
,
$$
and $h_{\rm ex}$ is the exchange field.
At some point, it will be convenient to assume
that the phase takes the particular value
$\varphi^{(\sigma)}=0$.
With this special value of the phase,
the form of the 3D Green's function (\ref{eq:spectral-ferro})
is identical to the
effective 
Green's function (\ref{eq:single-Ferro}).
The coherence length
in ferromagnetic metals is much shorter than
in usual metals so that ferromagnetism
is often treated in a semi-classical description
(see~\cite{Valet-Fert,Gijs-Bauer,Melin-Denaro}).
For instance,
the absence of Aharonov-Bohm oscillations 
reported in Ref.~\cite{Giroud} shows that
the coherence length in Co is smaller
than $0.3 \mu m$.
This can be incorporated
in our model by considering that the ``dissipation''
$\eta$ is not a small parameter. This results in
a finite coherence length, which is spin-dependent,
and inverse proportional
to the strength of dissipation:
$$
l_\phi^{(\sigma)} = \frac{1}{\eta}
\frac{ \hbar}{\sqrt{2 m}}
\sqrt{\epsilon_F+\sigma h_{\rm ex}+\omega}
.
$$
This simple phenomenological model contains
the relevant physics associated to ferromagnetic
metals. For instance, the phase coherence
length of spin-up electrons is larger than
the spin-down coherence length. The Green's function
(\ref{eq:spectral-ferro}) is infinite
when $\vx_i=\vx_j$,
which is also the case for the
superconductor Green's function (\ref{eq:spectral-3D}).
Local quantities can be obtained by using
$| \vx_i - \vx_j |=a_0$ instead of $\vx_i=\vx_j$.
With this condition, the local density of states
of the ferromagnet is given by
$$
\rho^{(\sigma)} = {1 \over 2 \pi^2}
\frac{m a_0^2}{\hbar^2}
\exp{ \left(- \frac{a_0}{2 l_\phi^{(\sigma)}} \right)}
.
$$
Spin-up electrons have thus a larger
density of states than spin-down electrons.

%%%%%%%%%%%%%%%%% FIGURE %%%%%%%%%%%%%%%%%%%%%%%%%
\begin{figure}[thb]
\centerline{\fig{8cm}{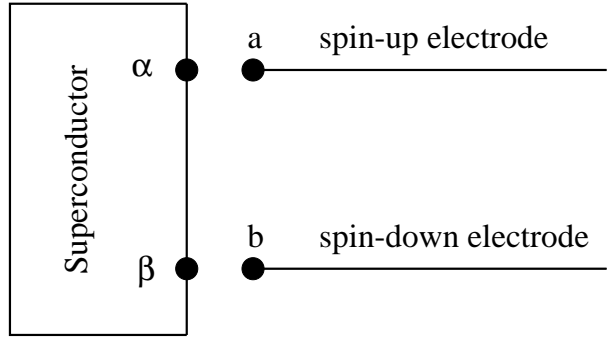}} 
\medskip
\caption{Schematic representation of the model
considered in section~\ref{sec:2channel}. Two
ferromagnetic electrodes are in contact
with a superconductor. A voltage $V_S$
is applied on the superconductor while 
the voltages $V_a$ and $V_b$ are
applied on the ferromagnetic electrodes.} 
\label{fig:schema1}
\end{figure}
%%%%%%%%%%%%%%%%%%%%%%%%%%%%%%%%%%%%%%%%%%%%%%%%%%

%%%%%%%%%%%%%%%%%%%%%%%%%%%%%%%%%%%%%%%%%%%%%%%%%%%%%%%%%%%%%
\section{Single channel electrodes: (I) Two electrodes
with $100\%$ spin polarization}
\label{sec:two-channel}
\label{sec:2channel}
\label{sec:tr-corr-pair}
In this section, we consider a model in which
a single channel spin-up electrode $a$ and a 
single channel spin-down
electrode $b$ are in contact with a superconductor.
We assume in sections~\ref{sec:4.1}
and~\ref{sec:4.2}
that $a$ is a spin-up channel and $b$
is a spin-down channel. The result for
parallel spin orientations is given
in section~\ref{sec:4.3}.

\subsection{Derivation of the transport formula}
\label{sec:4.1}
\subsubsection{Solution of the Dyson equation}

Let us first calculate the Nambu representation
of the propagators. The starting point is the
chain of Dyson equations given by (\ref{eq:Dy1}):
\begin{equation}
\label{eq:chain}
\left[ \begin{array}{c}
\HG^{a,a} \\ \HG^{b,a}
\end{array}
\right] =
\left[ \begin{array}{c} g^{a,a} \\ 0 \end{array}
\right]
+ \left[ \begin{array}{cc}
\HK^{a,a} & \HK^{a,b} \\
\HK^{b,a} & \HK^{b,b}
\end{array} \right]
\left[ \begin{array}{c}
\HG^{a,a} \\ \HG^{b,a}
\end{array} \right]
,
\end{equation}
where we used the notation
$\HK^{a_i,a_j} =
\Hg^{a_i,a_i} \Ht^{a_i,\alpha_i}
\Hg^{\alpha_i,\alpha_i'}
\Ht^{\alpha_i',a_i'}
$.
The solution of Eq.~\ref{eq:chain} is
\begin{equation}
\label{eq:block-G}
\left[ \begin{array}{cc}
G^{a,a}_{1,1} & G^{a,b}_{1,2} \\
G^{b,a}_{2,1} & G^{b,b}_{2,2}
\end{array} \right]
= \frac{1}{{\cal D}_{\rm AF}}
\left[ \begin{array}{cc}
g^{a,a}_{1,1} \left[ 1 - |t^{b,\beta}|^2
g_{2,2}^{b,b} g^{\beta,\beta} \right] &
-g^{b,b}_{2,2} t^{a,\alpha}
t^{b,\beta} g^{a,a}_{1,1}
f^{\alpha,\beta} \\
-g_{a,a}^{1,1} t^{a,\alpha}
t^{b,\beta}
g^{b,b}_{2,2} f^{\beta,\alpha}
& g^{b,b}_{2,2}
\left[ 1 - |t^{a,\alpha}|^2
g^{a,a}_{1,1} g^{\alpha,\alpha} \right]
\end{array}
\right]
,
\end{equation}
where the determinant ${\cal D}_{\rm AF}$ is
given by
\begin{equation}
\label{eq:D-AF-1}
{\cal D}_{\rm AF} = \left[ 1 - |t^{a,\alpha}|^2
g^{a,a}_{1,1} g^{\alpha,\alpha} \right]
\left[ 1 - |t^{b,\beta}|^2
g^{b,b}_{2,2} g^{\beta,\beta}
\right]
- |t^{a,\alpha}|^2
|t^{b,\beta}|^2
g^{a,a}_{1,1} g^{b,b}_{2,2}
f^{\alpha,\beta} f^{\beta,\alpha}
,
\end{equation}
and $g$ and $f$ have been defined as the components
of the Nambu matrix in (\ref{eq:fg}) and
(\ref{eq:spectral-3D}).
If not specified, all Green's functions in a given
formula stand as well for advanced and retarded
functions, and similarly for the determinant
${\cal D}_{\rm AF}$ given by Eq.~(\ref{eq:D-AF-1})
and the determinant ${\cal D}_{\rm F}$ that will
be introduced latter.
The matrix in 
Eq.~\ref{eq:block-G} contains the non vanishing
Nambu components of the renormalized propagator.
Because we assume a complete spin polarization,
the other Nambu components are vanishing.
For instance 
$G^{a,a}_{1,2}=G^{a,a}_{2,1}=G^{a,a}_{2,2}=0$.

\subsection{Exact expression of the current}
\label{sec:4.2}
Using the expression of the Keldysh propagator
(see Appendix~\ref{sec:app:keldysh}),
we deduce the final expression of the 
spin-up current in electrode $a$:
\begin{eqnarray}
\label{eq:courant-2contact}
\label{eq:2c-qp}
I_{1,1}^{a,\alpha} &=& - 4 \pi^2 |t_{a,\alpha}|^2
\int d \omega
\left[ n_F(\omega-\mu_a) - n_F(\omega-\mu_S) \right]
\rho^{a,a}_{1,1} \rho_g^{\alpha,\alpha}\\\nonb
&&\times
\frac{1}{\DA \DR}
\left[ 1 - |t_{b,\beta}|^2
g^{b,b,A}_{2,2}
g^{\beta,\beta,A} \right]
\left[ 1 - |t_{b,\beta}|^2
g^{b,b,R}_{2,2}
g^{\beta,\beta,R} \right]\\
\label{eq:2c-mixed1}
&+& 2 i \pi |t_{a,\alpha}|^2
|t_{b,\beta}|^2
\int d \omega
\left[ n_F(\omega-\mu_a) - n_F(\omega-\mu_S) \right]
\rho^{a,a}_{1,1}g_{2,2}^{b,b,A}
\\\nonb&&\times \frac{1}{\DA \DR}
f^{\alpha,\beta,A}
f^{\beta,\alpha,A} 
\left[ 1 - |t_{b,\beta}|^2
g^{b,b,R}_{2,2}
g^{\beta,\beta,R} \right]\\
\label{eq:2c-mixed2}
&-& 2 i \pi |t_{a,\alpha}|^2
|t_{b,\beta}|^2
\int d \omega
\left[ n_F(\omega-\mu_a) - n_F(\omega-\mu_S) \right]
\rho^{a,a}_{1,1}
g_{2,2}^{b,b,R} \\\nonb&&\times \frac{1}{\DA \DR}
 f^{\alpha,\beta,R}
f^{\beta,\alpha,R} 
\left[ 1 - |t_{b,\beta}|^2
g^{b,b,A}_{2,2}
g^{\beta,\beta,A} \right]\\
\label{eq:2c-andreev}
&-& 4 \pi^2 |t_{a,\alpha}|^2
|t_{b,\beta}|^2
\int d \omega
\left[ n_F(\omega-\mu_{b}) - n_F(\omega-\mu_S) \right]
\frac{1}{\DA \DR}
\rho_{1,1}^{a,a}
\rho_{2,2}^{b,b}
f^{\alpha,\beta,R} f^{\beta,\alpha,A}
,
\end{eqnarray}
which generalizes the result obtained by
Cuevas {\sl et al.} in Ref.~\cite{Cuevas}
in the case of a single conduction channel.
From the density of state prefactors, we see that
there are two type of contributions to the current:
(i) {\sl The quasiparticle current}, which is proportional
to the product of the density of state in the
superconductor ($\rho_g$) and one of the ferromagnetic electrodes
(for instance $\rho_a$);
(ii) {\sl The crossed Andreev current} which is proportional
to the product $\rho_a \rho_b$ of the density of state in the
two ferromagnetic electrodes. 
The term (\ref{eq:2c-qp}) contributes only
to quasiparticle current. The term (\ref{eq:2c-andreev})
contributes only to Andreev reflection. The mixed
terms (\ref{eq:2c-mixed1})~--~(\ref{eq:2c-mixed2})
contribute both to the quasiparticle and
Andreev current.

%%%%%%%%%%%%%%%%%%%%%%%%%%%%%%%%%%%%%%%%%%%%%%%%%%%%%%%%%%%%%

\subsection{Two-terminal conductance matrix}
\label{sec:4.3}
To understand the meaning of the transport
formula (\ref{eq:courant-2contact}), it is useful to
describe transport
across the multiterminal structure
in terms of a
differential conductance matrix:
\begin{equation}
\hat{\cal G} =
\left[ \begin{array}{cc}
\CG_{a,a} & \CG_{a,b} \\
\CG_{b,a} & \CG_{b,b} \end{array}\right]
,
\label{eq:current-mat-2}
\end{equation}
where the matrix elements are given by
\begin{equation}
\label{eq:def-mat-elem}
\CG_{a_i,a_j}(V_a,V_b) =
\frac{ \partial I_{a_i}}{\partial V_{a_j}}
(V_a,V_b)
.
\end{equation}
The conductance matrix (\ref{eq:current-mat-2})
encodes all information about
transport in the two-terminal structure.
The off-diagonal matrix elements should
satisfy a symmetry relation:
${\cal G}_{a,b}(V_a,V_b) =
{\cal G}_{b,a}(V_b,V_a)$.
If the electrodes
have an antiparallel spin orientation, subgap
current is transported by Cooper pairs
if $\omega<\Delta$, in which case we have
$I_a=I_b$.
This implies
an additional symmetry relation:
${\cal G}_{a,a}(V_a,V_b) =
{\cal G}_{b,a}(V_a,V_b)$,
and ${\cal G}_{b,b}(V_a,V_b) =
{\cal G}_{a,b}(V_a,V_b)$.
If the electrodes have
a parallel spin orientation, subgap current
is due to elastic cotunneling, in which
case $I_a=-I_b$. The additional
symmetry relation reads ${\cal G}_{a,a}(V_a,V_b) =
- {\cal G}_{b,a}(V_a,V_b)$, and
${\cal G}_{b,b}(V_a,V_b) =
- {\cal G}_{a,b}(V_a,V_b)$.

\subsubsection{Sub-gap conductance matrix:
effective Green's functions}
\label{sec:all-below-2chan}
\label{sec:1site-below}
In this section as well as in section~\ref{sec:1site-above},
we assume that $\cos{\varphi}=0$ so that we can use
effective Green's functions to evaluate the transport formula
and work out the basic physics on simple grounds.
The validity of this assumption will be discussed in
section~\ref{sec:comp-2}.

%%%%%%%%%%%%%%%%% FIGURE %%%%%%%%%%%%%%%%%%%%%%%%%
\begin{figure}[thb]
\centerline{\fig{4cm}{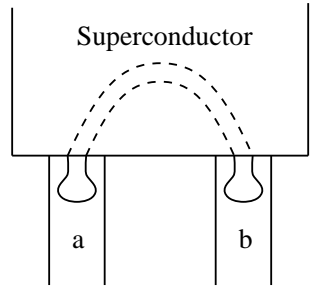}} 
\medskip
\caption{The diagram associated
to Andreev reflection in the two-channel model.
\base
} 
\label{fig:d1}
\end{figure}
%%%%%%%%%%%%%%%%%%%%%%%%%%%%%%%%%%%%%%%%%%%%%%%%%%

\paragraph{Antiparallel magnetizations:}
The sub-gap
current of the two-channel model with antiparallel
magnetizations
originates from the
non local Andreev reflections
which are shown schematically on Fig.~\ref{fig:d1}.
This can be seen by
inserting the effective
Green's functions into the
transport formula:
\begin{eqnarray}
\label{eq:sub-gap-1}
{\cal G}_{a,a} &=& {\cal G}_{b,a} =
- \frac{4 \Gamma_a \Gamma_b}{
| {\cal D}_{\rm AF}(\omega=V_a) |^2}
\left[ f^{\alpha,\beta}
(\omega=V_a) \right]^2\\
{\cal G}_{a,b} &=& {\cal G}_{b,b} =
- \frac{4 \Gamma_a \Gamma_b}{
| {\cal D}_{\rm AF}(\omega=V_b) |^2}
\left[ f^{\alpha,\beta}
(\omega=V_b) \right]^2
\label{eq:sub-gap-2}
,
\end{eqnarray}
where $\Gamma_a = \pi |t_{a,\alpha}|^2
\rho_a$ is the spectral line-width
associated to electrode $a$, and a similar
expression holds for $\Gamma_b$.
We used the fact that $g^{\alpha,\beta}$
and $f^{\alpha,\beta}$ are real numbers
below the superconducting gap.
The expression of ${\cal D}_{\rm AF}$ is the following:
\begin{equation}
\label{eq:D-below}
|{\cal D}_{\rm AF}(\omega)|^2 =
\left\{ 1 - \Gamma^a \Gamma^b \left[
g^{\alpha,\alpha} g^{\beta,\beta}
- ( f^{\alpha,\beta} )^2 \right]
\right\}^2
+ \left( \Gamma^a g^{\alpha,\alpha}
+ \Gamma^b g^{\beta,\beta} \right)^2
.
\end{equation}

\paragraph{Parallel magnetizations:}
The same calculation can be done if the electrodes
have a parallel spin orientation. We find
\begin{eqnarray}
\label{eq:sub-gap-bis-1}
{\cal G}_{a,a} &=&- {\cal G}_{b,a} =
- \frac{4 \Gamma_a \Gamma_b}{
| {\cal D}_{\rm F}(\omega=V_a) |^2}
\left[ g^{\alpha,\beta}
(\omega=V_a) \right]^2\\
- {\cal G}_{a,b} &=& {\cal G}_{b,b} =
- \frac{4 \Gamma_a \Gamma_b}{
| {\cal D}_{\rm F}(\omega=V_b) |^2}
\left[ g^{\alpha,\beta}
(\omega=V_b) \right]^2
\label{eq:sub-gap-bis-2}
,
\end{eqnarray}
with
\begin{equation}
\label{eq:D-below-F}
|{\cal D}_{\rm F}(\omega)|^2 =
\left\{ 1 - \Gamma^a \Gamma^b \left[
g^{\alpha,\alpha} g^{\beta,\beta}
- ( g^{\alpha,\beta} )^2 \right]
\right\}^2
+ \left( \Gamma^a g^{\alpha,\alpha}
+ \Gamma^b g^{\beta,\beta} \right)^2
.
\end{equation}
There are two differences
between the situations with antiparallel and parallel
spin orientations. First, the Andreev reflection
transport with antiparallel spin orientations
is controlled by
the anomalous propagator $f^{\alpha,\beta}$
while the elastic cotunneling transport
with parallel spin orientations
is controlled by the ordinary propagator
$g^{\alpha,\beta}$.
The second difference is in the sign of the 
off-diagonal conductance matrix elements.
The four matrix elements have the same sign
in the case of Andreev reflection because
transport is mediated by Cooper pairs.
The off-diagonal matrix elements have a 
sign opposite to the diagonal matrix elements
in the case of elastic cotunneling because
transport is due to single electron tunneling
between the two electrodes.

\subsubsection{Conductance matrix above the
superconducting gap: effective Green's functions}
\label{sec:all-above-2chan}
\label{sec:1site-above}
Let us now assume that the voltage $V_a$ is above
the superconducting gap and that the electrodes
have an antiparallel spin orientation. With
the notation
$g^{\alpha,\beta,A,R} = \pm
i |g^{\alpha,\beta}|$, and
$f^{\alpha,\beta,A,R} = \pm
i |f^{\alpha,\beta}|$,
the extra diagonal terms of the
conductance matrix take the form
$$
\CG_{b,a} = -
\frac{4 \Gamma_a \Gamma_b}{\D_{\rm AF}^2}
|f^{\alpha,\beta}|^2
,
$$
which should be evaluated at the energy
$\omega=V_a$.
The diagonal conductance matrix element is
the sum of a crossed contribution and a quasiparticle
contribution:
$\CG_{a,a} =
- \CG_{b,a} -
\CG_{a,a}^{qp}$. The quasiparticle contribution 
is the sum of a direct and a crossed term:
\begin{equation}
\label{eq:G-aa-2-chan}
\CG_{a,a}^{qp} =  \frac{4 \pi
\Gamma_a \rho_g}{\D_{\rm AF}^2} \left[
\left( 1 + \Gamma_b |g^{\beta,\beta}|
\right)^2 - \Gamma_b^2 |
f^{\alpha,\beta} |^2 \right]\\\nonb
,
\end{equation}
where the denominator
${\cal D}_{\rm AF}$ is a real number:
$$
\D_{\rm AF} =\DA_{\rm AF}=\DR_{\rm AF}
= 1 + \Gamma^a \Gamma^b \left[
|g^{\alpha,\alpha} g^{\beta,\beta}|
- |f^{\alpha,\beta}|^2 \right]
+ \Gamma^a |g^{\alpha,\alpha}|
+ \Gamma^b |g^{\beta,\beta}|
.
$$
In the limit $|\vx_a - \vx_b| \rightarrow
+ \infty$ in which the separation between
the contacts becomes very large,
the extra diagonal conductance matrix
elements are vanishingly small.
The quasiparticle term 
reduces to the conductance of a single channel
metal~--~metal contact:
$$
\CG_{a,a}^{qp} = \frac{4 \pi \Gamma_a \rho_g}
{ \left( 1 + \Gamma_a |g^{\alpha,\alpha}| \right)^2}
.
$$

\subsection{Phase resolved {\sl versus}
averaged conductance}
\label{sec:comp-2}
Given the form (\ref{eq:spectral-3D}) of the
superconductor Green's function,
we see that the conductance depends explicitly
on the electronic phase difference $\varphi=k_F |\vx_a - \vx_b|$.
This leads us to calculate the conductance in
two different ways:
\begin{itemize}
\item[(i)] The phase-resolved conductance
${\cal G}(\varphi)$. We will focus more
especially on the case $\varphi=\pi/2$.
For this special value of the phase difference,
the Green's function of the superconductor 
coincides with the effective Green's function
in the limit $|\vx_a - \vx_b| \rightarrow
a_0$.
\item[(ii)] The averaged conductance
\begin{equation}
\label{eq:av-cond}
\LL {\cal G}(\varphi) \RR =
\frac{1}{2 \pi} \int_0^{2 \pi} {\cal G}(\varphi)
d \varphi
.
\end{equation}
This phase averaging is used to mimic the physics
of extended contacts that will be considered
later in section~\ref{sec:multi-chan}.
\end{itemize}

\begin{figure}[thb]
\centerline{\fig{8cm}{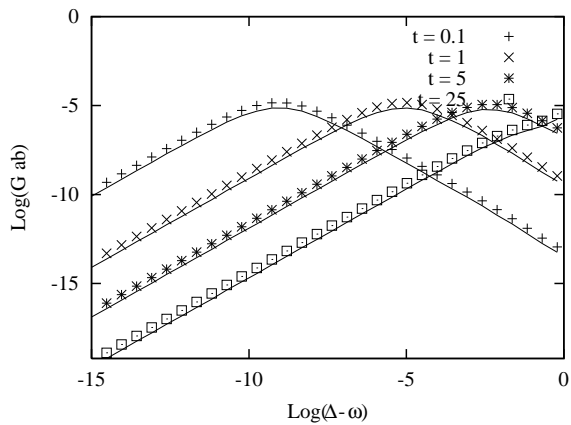}} 
\medskip
\caption{Variation of the logarithm of the
crossed conductance
$\log{\cal G}_{a,b}$ {\sl versus}
$\log(\Delta-\omega)$ ($\omega<\Delta$).
${\cal G}_{a,b}$ is in units of $e^2/h$.
The points
correspond to the phase-resolved conductance
with $\varphi=\pi/2$, namely, to the
generalized effective Green's
functions. The solid lines
correspond to the average conductance
(\ref{eq:av-cond}).
We used the parameters
$m=0.01$, $k_F=1$,
$\epsilon_F=50$, $a_0=1$, $\Delta=1$.
The distance between the contacts
is $D=100$ and the superconductor coherence
length is $\xi_0=\epsilon_F / (k_F \Delta)=50$.
\base
} 
\label{fig:A}
\end{figure}

To determine the role played by phase averaging, we
compare the phase-dependent and average conductances
(see the end of section~\ref{sec:3.1.3}).
It is visible on Fig.~\ref{fig:A} that the
effective Green's function conductance ({\sl i.e.} with
$\varphi=\pi/2$) follows closely the average
conductance. This shows that the effective Green's function
conductance contains already the relevant physics,
as far as the two-channel problem is concerned.

%%%%%%%%%%%%%%%%% FIGURE %%%%%%%%%%%%%%%%%%%%%%%%%
%%%%%%%%%%%%%%%%%%%%%%%%%%%%%%%%%%%%%%%%%%%%%%%%%%

It is also visible on Fig.~\ref{fig:A} that there
is a cross-over energy $\omega^*$
(thus a cross-over voltage $V^*=
\omega^*/e$)
at which the crossed conductance reaches a maximum.
If $\omega<\omega^*$ the crossed conductance behaves
like ${\cal G}_{a,b}
\sim 1 / (\Delta - \omega)$.  
If $\omega^*<\omega<\Delta$, the crossed
conductance behaves like ${\cal G}_{a,b}
\sim \Delta - \omega$. Only when
$\omega < \omega^*$ does our approach
coincide with the lowest
order tunnel perturbation theory.
The analysis based on generalized effective
Green's functions is
simplified if one assumes that $D \gg a_0$,
but $D$ can be small or large compared to the
superconducting coherence length $\xi(\omega)$
(see Eq.~\ref{eq:xi-om}).
The behavior in the energy
range $\omega^*<\omega<\Delta$ is non perturbative,
and can be understood 
by retaining in ${\cal D}_{\rm AF}$ only the leading
divergence, which is generated by the quartic terms
(see Eq.~\ref{eq:D-below}):
$$
| {\cal D}_{\rm AF}|^2 \simeq \Gamma^4 \left[
g^{\alpha,\alpha} g^{\beta,\beta}
- \left( f^{\alpha,\beta} \right)^2 \right]^2
,
$$
from what we deduce the expression of the 
crossed conductance
\begin{equation}
\label{eq:cond-sup}
{\cal G}_{a,b} \simeq {4 \over \Gamma^2}
\left( \frac{a_0}{D} \right)^2
\left( \frac{2 \pi \hbar^2}{m a_0^2} \right)^2
\left(\frac{\Delta^2 - \omega^2}{\Delta^2} \right)
\exp{\left(-{D \over \xi(\omega)}\right)}
,
\end{equation}
valid in 
the energy range $\omega > \omega^*$.
The
expression of the cross-over energy
$\omega^*$ is obtained by equating
the quadratic and quartic terms in ${\cal D}_{\rm AF}$.
This leads to
\begin{equation}
\label{eq:omega-*}
\omega^* = \Delta \sqrt{1 - \left(
\Gamma \frac{m a_0^2}{2 \pi \hbar^2} \right)^2}
.
\end{equation}
To obtain the conductance in the energy range $\omega<\omega^*$,
we  expand ${\cal D}_{\rm AF}$ up to order $\Gamma^2$:
$$
|{\cal D}_{\rm AF}|^2 \simeq 1 + 2 \Gamma^2
\left( g^{\rm loc} \right)^2
+2 \Gamma^2 \left( f^{\alpha,\beta} \right)^2
,
$$
where $g^{\rm loc} = g^{\alpha,\alpha}
=g^{\beta,\beta}$ denotes the local propagator
in the superconductor.
One can show easily that $f^{\alpha,\beta} \ll g^{\rm loc}$
because $D \gg a_0$,
from what we deduce
\begin{equation}
|{\cal D}_{\rm AF}|^2 \simeq 1 + 2 \Gamma^2
\left( \frac{m a_0^2}{2 \pi \hbar^2} \right)^2
\frac{ \omega^2}{\Delta^2 - \omega^2}
.
\label{eq:D-approx}
\end{equation}
${\cal D}_{\rm AF}$ is close to unity only when
$\omega < \omega_0$, with
\begin{equation}
\label{eq:omega-0}
\omega_0 = \frac{ \Delta}
{ \sqrt{ 1 + 2 \Gamma^2 \left(
\frac{ m a_0^2}{2 \pi \hbar^2} \right)^2}}
.
\end{equation}
Comparing (\ref{eq:omega-*}) and
(\ref{eq:omega-0}), we see that
$\omega^*$ is larger than $\omega_0$ but $\omega_0$ and
$\omega^*$ have the same order of magnitude
is $\Gamma$ is small (which is the case
on Fig.~\ref{fig:A}). We deduce that
${\cal D}_{\rm AF}=1$ in
the relevant energy range $\omega < \omega_0 < \omega^*$.
From what we obtain the
conductance in the energy range $\omega<\omega_0$:
\begin{equation}
\label{eq:cond-inf}
{\cal G}_{a,b} = 4 \Gamma^2 \left( \frac{a_0}{D} \right)^2
\left( \frac{ m a_0^2}{2 \pi \hbar^2} \right)^2
\frac{ \Delta^2 }{\Delta^2 - \omega^2}
\exp{\left( - \frac{D}{\xi(\omega)} \right)}
,
\end{equation}
identical to the one obtained in lowest order
perturbation theory.
Evaluating (\ref{eq:cond-sup}) and
(\ref{eq:cond-inf}) at $\omega=\omega^*$
or $\omega=\omega_0$ leads to
the maximal value of the conductance:
\begin{equation}
\label{eq:G-max}
{\cal G}_{a,b}^{\rm max} \simeq \frac{e^2}{h}
\left( \frac{a_0}{D} \right)^2 \exp{\left(
- \frac{D}{\xi(\omega^*)} \right)}
,
\end{equation}
where the numerical prefactor of order unity
cannot be obtained from this simple estimate.

It is well known from the BTK 
scattering approach~\cite{BTK}
that the conductance of a normal metal~--
superconductor contact is equal to twice the quantum
of conductance
$e^2/h$ per spin channel if $\omega=\Delta$,
{\sl regardless the value of the interface scattering}.
The same behavior occurs in the Keldysh
formalism treatment by Cuevas {\sl et al.}~\cite{Cuevas}.
This type of resonance can be
properly described only with a non perturbative
approach.
Eq.~\ref{eq:G-max} constitutes a generalization
of the BTK behavior to the case of spatially
separated contacts having a phase
difference $\varphi=\pi/2$. Given
that the average and phase-resolved conductances
follow closely each other (see Fig.~\ref{fig:A}),
it is expected that (\ref{eq:G-max}) is also valid
for the average conductance,
but with an extra reduction factor.

%%%%%%%%%%%%%%%%%%%%%%%%%%%%%%%%%%%%%%%%%%%%%%%%%%%%%%%%
\section{Single channel electrodes: (II) General solution}
\label{sec:general}

%%%%%%%%%%%%%%%%% FIGURE %%%%%%%%%%%%%%%%%%%%%%%%%
\begin{figure}[thb]
\centerline{\fig{6cm}{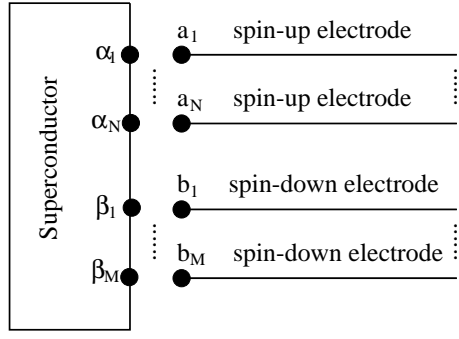}} 
\medskip
\caption{Schematic representation of the model
with $N$ ferromagnetic spin-up electrodes and $M$
ferromagnetic spin-down electrodes.
\base
} 
\label{fig:schema3}
\end{figure}
%%%%%%%%%%%%%%%%%%%%%%%%%%%%%%%%%%%%%%%%%%%%%%%%%%

Now we consider a model in which $N$ spin-up
electrodes and $M$ spin-down electrodes are
in contact with a superconductor (see
Fig.~\ref{fig:schema3}). All the necessary
details can be found in Appendix~\ref{app:B}.
The final form of the current is
\begin{eqnarray}
\label{eq:transport-gene}
\label{eq:tr-gen-qp}
I^{\alpha_1,a_1} &=&-4 \pi^2 |t^{a_1,\alpha_1}|^2
\int d \omega
\left[ n_F(\omega-\mu_{a_1})
- n_F(\omega-\mu_S) \right]
\rho^{\alpha_1,\alpha_1}
\rho_g^{\alpha_1,\alpha_1}
\frac{1}{\DA \DR}
\tilde{\cal M}_{a_1,a_1}^A
\tilde{\cal M}_{a_1,a_1}^R\\
\label{eq:tr-gen-mixed1}
&-&2 i \pi \sum_{k=2}^N
(-)^{k+1}
t^{a_1,\alpha_1} t^{a_k,\alpha_k}
\int d \omega
\left[ n_F(\omega-\mu_{a_1})
- n_F(\omega-\mu_S) \right]\times\\&&
\nonb
\frac{1}{\DA \DR}
\left(
g^{\alpha_1,\alpha_k,R}
\tilde{\cal M}^R_{a_1,a_k}
\tilde{\cal M}^A_{a_1,a_1} -
g^{\alpha_1,\alpha_k,A}
\tilde{\cal M}^A_{a_1,a_k}
\tilde{\cal M}^R_{a_1,a_1} \right)\\
\label{eq:tr-gen-mixed2}
&-&2 i \pi \sum_{k=2}^N
(-)^{k+N}
t^{a_1,\alpha_1} t^{b_k,\beta_k}
\int_{\mu_{a_1}}^{\mu_\alpha}
d \omega
\left[ n_F(\omega-\mu_{a_1})
- n_F(\omega-\mu_S) \right]\times\\&&
\nonb\frac{1}{\DA \DR}
\left(
f^{\alpha_1,\beta_k,R}
\tilde{\cal M}^R_{a_1,b_k}
\tilde{\cal M}^A_{a_1,a_1} -
f^{\alpha_1,\beta_k,A}
\tilde{\cal M}^A_{a_1,b_k}
\tilde{\cal M}^R_{a_1,a_1} \right)\\
\label{eq:tr-gen-EC}
&+& 4 \pi^2 \sum_{k=2}^N
|t^{a_1,\alpha_1}|^2
|t^{a_k,\alpha_k}|^2
\int d \omega
\left[ n_F(\omega-\mu_{a_k})
- n_F(\omega-\mu_S) \right] \times\\
\nonb&&
\rho^{a_1,a_1}_{1,1}
\rho^{a_k,a_k}_{1,1}
\tilde{g}^{\alpha_1,\alpha_k,R}
\tilde{g}^{\alpha_k,\alpha_1,A}\\
\label{eq:tr-gen-AR}
&-& 4 \pi^2 \sum_{k=1}^M
|t^{a_1,\alpha_1}|^2
|t^{b_k,\beta_k}|^2
\int d \omega
\left[ n_F(\omega-\mu_{b_k})
- n_F(\omega-\mu_S) \right] \times\\
\nonb&&
\rho^{a_1,a_1}_{1,1}
\rho_{2,2}^{b_k,b_k}
\tilde{f}^{\alpha_1,\beta_k,R}
\tilde{f}^{\beta_k,\alpha_1,A}
,
\end{eqnarray}
where we used the notation
\begin{eqnarray}
\tilde{g}^{\alpha_i,\alpha_j}
&=& \frac{ \tilde{\cal M}_{a_j,a_i}}
{t^{a_i,\alpha_i}
t^{a_j,\alpha_j}
g^{a_i,a_i} {\cal D}}\\
\tilde{f}^{\alpha_i,\beta_j}
&=& \frac{ \tilde{\cal M}_{b_j,a_i}}
{t^{a_i,\alpha_i}
t^{b_j,\beta_j}
g^{a_i,a_i} {\cal D}}
\end{eqnarray}
for the renormalized propagators.
There are three types of processes
involved in the transport formula:
(i) {\sl The quasiparticle term} (\ref{eq:tr-gen-qp})
which is proportional
to $\rho^{\alpha_1,\alpha_1} \rho_g$;
(ii) {\sl The elastic cotunneling term} (\ref{eq:tr-gen-EC})
in which spin-up electrons from electrode
$a_k$ are transfered into electrode $1$.
The elastic cotunneling terms are proportional to
$\rho_{1,1}^{a_1,a_1} \rho_{1,1}^{a_k,a_k}$
(iii) {\sl The Andreev reflection term} (\ref{eq:tr-gen-AR})
which are proportional to 
$\rho_{1,1}^{a_1,a_1} \rho_{2,2}^{b_k,b_k}$.
The mixed terms (\ref{eq:tr-gen-mixed1}) and
(\ref{eq:tr-gen-mixed2}) contribute the
three types of processes.

%%%%%%%%%%%%%%%%%%%%%%%%%%%%%%%%%%%%%%%%%%%%%%%%%%%%%%%%

\section{Single channel electrodes: (III) Three electrodes
with $100\%$ spin polarization}
\label{sec:3channel}

\begin{figure}[thb]
\centerline{\fig{6cm}{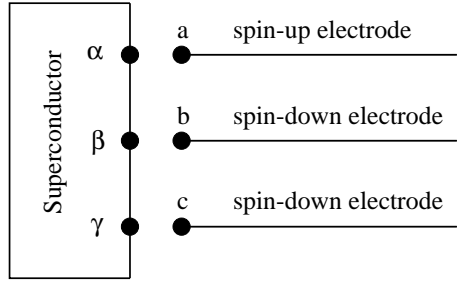}} 
\medskip
\caption{Schematic representation of the three channel
model
considered in section~\ref{sec:3channel}.
\base
} 
\label{fig:schema2}
\end{figure}

Let us now consider a three-terminal problem. We consider
that each of the three electrodes has $100\%$ spin polarization
(see Fig.~\ref{fig:schema2}).
The aim is to have Andreev and cotunneling processes
occurring in the same multiterminal device, which
allows a direct comparison of these two basic
processes.
The transport formula can be deduced easily from
the general solution obtained in section~\ref{sec:general}.

%%%%%%%%%%%%%%%%% FIGURE %%%%%%%%%%%%%%%%%%%%%%%%%
%%%%%%%%%%%%%%%%%%%%%%%%%%%%%%%%%%%%%%%%%%%%%%%%%%

\subsection{Three-terminal conductance matrix}
The three-terminal conductance matrix generalizing
(\ref{eq:current-mat-2})
takes the form
\begin{equation}
\hat{\cal G} =
\left[ \begin{array}{ccc}
\CG_{a,a} & \CG_{a,b} & \CG_{a,c} \\
\CG_{b,a} & \CG_{b,b} & \CG_{b,c} \\
\CG_{c,a} & \CG_{c,b} & \CG_{c,c} 
\end{array} \right]
.
\end{equation}
Let us assume that
electrode $a$ has
a spin-up magnetization, and electrodes $b$
and $c$ have a spin-down magnetization (see Fig.~\ref{fig:schema2}).
We use the same procedure as in section~\ref{sec:1site-below}
to obtain the conductance matrix elements.
Namely, we assume that $\cos{\varphi}=0$ and
replace the Green's functions by effective
Green's functions.

\subsubsection{Conductance matrix below the superconducting gap:
effective Green's functions}
Let us first give the form of the off diagonal
matrix elements:
\begin{eqnarray}
\label{eq:off-1}
\CG_{a,b} &=& \CG_{b,a} = - \frac{4 \Gamma_a \Gamma_b}
{\DA \DR} \left[ \left( f^{\alpha,\beta}
\right)^2 + \Gamma_c^2 \left( f^{\alpha,\gamma}
g^{\beta,\gamma} - f^{\alpha,\beta}
g^{\gamma,\gamma} \right)^2 \right] \\
\label{eq:off-2}
\CG_{a,c} &=&\CG_{c,a} =- \frac{4 \Gamma_a \Gamma_c}
{\DA \DR} \left[ \left( f^{\alpha,\gamma}
\right)^2 + \Gamma_b^2 \left( f^{\alpha,\beta}
g^{\beta,\gamma} - f^{\alpha,\gamma}
g^{\beta,\beta} \right)^2 \right] \\
\label{eq:off-3}
\CG_{b,c} &=&\CG_{c,b}  =\frac{4 \Gamma_b \Gamma_c}
{\DA \DR} \left[ \left( g^{\beta,\gamma}
\right)^2 + \Gamma_a^2 \left( f^{\alpha,\beta}
f^{\alpha,\gamma} - g^{\alpha,\alpha}
g^{\beta,\gamma} \right)^2 \right]
.
\end{eqnarray}
From the signs of these matrix elements,
and from the type of propagator involved,
we see that $\CG_{a,b}$ and $\CG_{a,c}$
correspond to Andreev reflection
while $\CG_{b,c}$ corresponds
to elastic cotunneling.

%%%%%%%%%%%%%%%%% FIGURE %%%%%%%%%%%%%%%%%%%%%%%%%
\begin{figure}[thb]
\centerline{\fig{8cm}{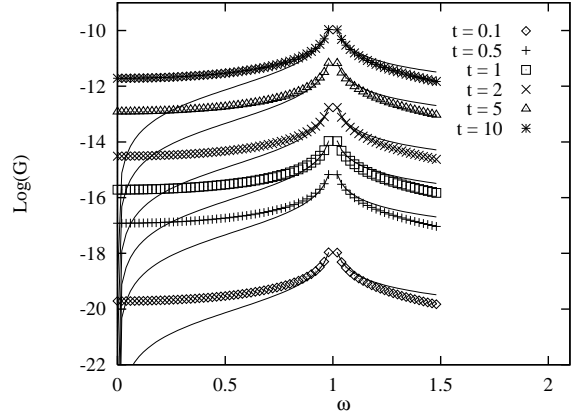}} 
\medskip
\caption{Variation of the logarithm of the
Andreev reflection conductance ${\cal G}_{a,c}$
and elastic cotunneling
conductance ${\cal G}_{b,c}$ {\sl versus} reduced
energy $\omega/\Delta$.
The points correspond to Andreev reflection
and the solid lines correspond to elastic cotunneling.
We have assumed that site $\alpha$ coincides
with site $\beta$ (see Fig.~\ref{fig:schema2}).
We used effective Green's functions with
$\varphi=\pi/2$.
The distance between the contacts
is $D=100$. On purpose, we did not show
the behavior for $\omega \simeq \Delta$
(see Fig.~\ref{fig:C}).
The parameters are identical as on Fig.~\ref{fig:A}.
\base
} 
\label{fig:B}
\end{figure}
%%%%%%%%%%%%%%%%%%%%%%%%%%%%%%%%%%%%%%%%%%%%%%%%%%

\subsubsection{Conductance matrix above the superconducting gap:
effective Green's functions}
Using the notation $f^{A,R} = \pm i |f|$, and
$g^{A,R} = \pm i |g|$,
we obtain the conductance matrix elements above
the superconducting gap:
\begin{eqnarray}
\CG_{a,b} &=& \CG_{b,a} = 
- \frac{4 \Gamma_a \Gamma_b}
{{\cal D}^A{\cal D}^R} \left[ |f^{\alpha,\beta}| +
\Gamma_c \left( |f^{\alpha,\beta}
g^{\gamma,\gamma}| - |f^{\alpha,\gamma}
g^{\beta,\gamma}| \right) \right]^2\\
\CG_{a,c} &=& \CG_{c,a}= - \frac{4 \Gamma_a \Gamma_c}
{{\cal D}^A{\cal D}^R} \left[ |f^{\alpha,\gamma}| +
\Gamma_b \left( |f^{\alpha,\gamma}
g^{\beta,\beta}| - |f^{\alpha,\beta}
g^{\beta,\gamma}| \right) \right]^2\\
\CG_{b,c} &=& \CG_{c,b}=- \frac{4 \Gamma_b \Gamma_c}
{{\cal D}^A{\cal D}^R} \left[ |g^{\beta,\gamma}| +
\Gamma_a \left( |g^{\beta,\gamma}
g^{\alpha,\alpha}| - |f^{\alpha,\beta}
f^{\alpha,\gamma}| \right) \right]^2
.
\end{eqnarray}
The diagonal coefficient $\CG_{a,a}$ takes
the form
\begin{eqnarray}
\CG_{a,a} &=& -\CG_{a,b} - \CG_{a,c}
+ \frac{4 \Gamma_a}{{\cal D}^A {\cal D}^R}
\left\{ -\pi \rho_g  \tilde{\cal M}^{a,a,A}
\tilde{\cal M}^{a,a,R}
\right.\\
&+& \Gamma_b^2 \left\{ |f^{\alpha,\beta}|^2
\left[ |g^{\beta,\beta}| + \Gamma_c \left(
2 + \Gamma_c |g^{\gamma,\gamma}| \right)
\left( |g^{\beta,\beta} g^{\gamma,\gamma}|
- |G^{\beta,\gamma}|^2 \right) \right] \right\}\\
&+& \Gamma_c^2 \left\{ |f^{\alpha,\gamma}|^2
\left[ |g^{\gamma,\gamma}| + \Gamma_b \left(
2 + \Gamma_b |g^{\beta,\beta}| \right)
\left( |g^{\beta,\beta} g^{\gamma,\gamma}|
- |g^{\beta,\gamma}|^2 \right) \right] \right\}\\
&+& \left. 2 \Gamma_b \Gamma_c
|f^{\alpha,\beta} f^{\alpha,\gamma} g^{\beta,\gamma}|
\left[ 1 - \Gamma_b \Gamma_c
\left( |g^{\beta,\beta} g^{\gamma,\gamma}|
- |g^{\beta,\gamma}|^2 \right) \right] \right\}
,
\end{eqnarray}
and similar expressions can be obtained
for $\CG_{b,b}$ and $\CG_{c,c}$. 

\subsection{Phase resolved {\sl versus} average
conductance}
Let us now
consider the special situation in which 
a ferromagnetic electrode with $100\%$ spin
polarization is at a distance $D$ away from
a normal metal electrode having
a zero spin polarization. Namely, we assume
that sites $\alpha$ and $\beta$ coincide
(see Fig.~\ref{fig:schema2}).
This provides
the simplest model containing
both Andreev reflection and elastic
cotunneling.

\begin{figure}[thb]
\centerline{\fig{8cm}{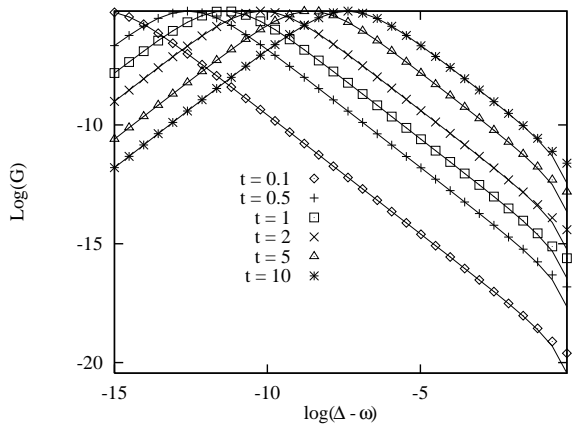}} 
\medskip
\caption{The same as Fig.~\ref{fig:B}
but for energies $\omega \simeq \Delta$.
\base
} 
\label{fig:C}
\end{figure}

Following section~\ref{sec:comp-2}, we evaluate
the phase-resolved conductance with
$\varphi_{a,b}=\varphi_{a,c}=
\varphi_{b,c}=\pi/2$. With this particular
value of the phases, the
phase-resolved conductance coincides
with the effective Green's function conductance.
We have shown on Fig.~\ref{fig:B}
the energy dependence of the 
Andreev reflection and elastic tunneling
conductances, evaluated with
$\varphi=\pi/2$. It is visible that
the Andreev reflection conductance
is larger than the elastic cotunneling
conductance by a factor $\Delta/\omega$.
The Andreev conductance
coincides with the elastic cotunneling conductance
only when
$\omega$ is close to $\Delta$ (see Fig.~\ref{fig:C}).
This behavior can be
understood from the
effective Green's function conductance
(Eqs.~\ref{eq:off-1}~--~\ref{eq:off-3}).

%%%%%%%%%%%%%%%%% FIGURE %%%%%%%%%%%%%%%%%%%%%%%%%
%%%%%%%%%%%%%%%%%%%%%%%%%%%%%%%%%%%%%%%%%%%%%%%%%%

%%%%%%%%%%%%%%%%% FIGURE %%%%%%%%%%%%%%%%%%%%%%%%%
\begin{figure}[thb]
\centerline{\fig{8cm}{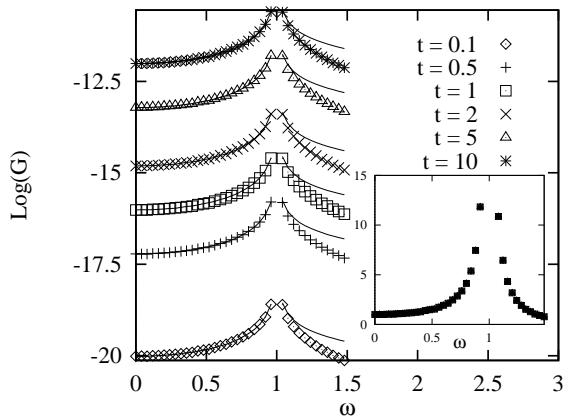}} 
\medskip
\caption{Variation of the logarithm of the
Andreev reflection (points) and elastic
cotunneling (solid lines)
conductances {\sl versus} energy
$\omega$. We averaged the conductance
over all possible phase configurations
(see Eq.~\ref{eq:average-3}).
The insert shows the
energy dependence of the rescaled conductance
${\cal G}(\omega) / {\cal G}(\omega=0)$.
The parameters are identical as on Fig.~\ref{fig:A}.
\base
} 
\label{fig:D}
\end{figure}
%%%%%%%%%%%%%%%%%%%%%%%%%%%%%%%%%%%%%%%%%%%%%%%%%%

To obtain the average conductance, we
assume that the phase variables are independent
random variables, and average the conductance
over all possible values of the phases:
\begin{equation}
\label{eq:average-3}
\LL {\cal G} \RR = 
\int_0^{2 \pi} {d \varphi_{a,b} \over 2 \pi}
\int_0^{2 \pi} {d \varphi_{a,c} \over 2 \pi}
\int_0^{2 \pi} {d \varphi_{b,c} \over 2 \pi}
{\cal G}(\varphi_{a,b},
\varphi_{a,c},\varphi_{b,c})
.
\end{equation}
It is visible on Fig.~\ref{fig:D} that
after phase averaging, the Andreev
reflection conductance coincides exactly
with the elastic cotunneling conductance
below the superconducting gap. For energies
close to $\Delta$, we find again the
existence of a maximum in the
conductance at an energy $\omega^*$.
The predictions of perturbation theory are
valid in the energy range $\omega<\omega^*$.
In this energy range,
all conductance spectra on Fig.~\ref{fig:D}
can be deduced from each other by
a simple rescaling
(see the insert of Fig.~\ref{fig:D}).

%%%%%%%%%%%%%%%%%%%%%%%%%%%%%%%%%%%%%%%%%%%%%%%%%%%%%%%%
\section{Multichannel electrodes}
\label{sec:multi-chan}

%%%%%%%%%%%%%%%%% FIGURE %%%%%%%%%%%%%%%%%%%%%%%%%
\begin{figure}[thb]
\centerline{\fig{8cm}{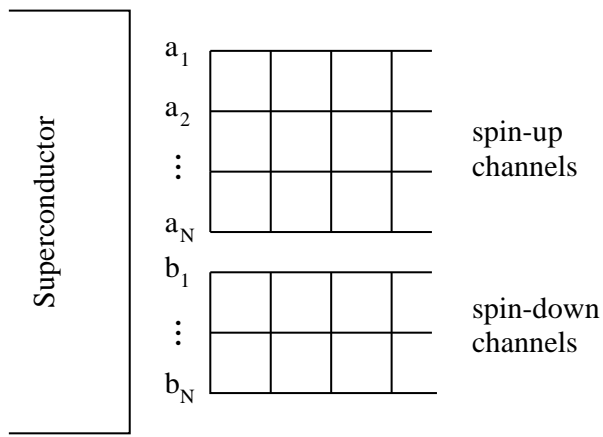}} 
\medskip
\caption{Schematic representation of the tight
binding model.
The contacts between the spin-up ferromagnet and
the superconductor are noted
$a_1$,..., $a_N$.
The contacts between the spin-down ferromagnet and
the superconductor are noted
$b_1$,..., $b_N$.
\base
} 
\label{fig:deJong}
\end{figure}
%%%%%%%%%%%%%%%%%%%%%%%%%%%%%%%%%%%%%%%%%%%%%%%%%%

\subsection{Transport formula}
We want to determine whether extended
contacts have a physics identical to the single
channel contacts considered in
sections~\ref{sec:two-channel},~\ref{sec:general}
and~\ref{sec:3channel}.
We are thus lead to investigate the
following situations:
\begin{itemize}
\item[(i)] {\sl The phase averaged conductance
of extended contacts}.
There are $N(N-1)/2$ phases $\varphi_{i,j}$
associated with a contact having $N$
channels. The phase-averaged conductance
is obtained by averaging the conductance
over all possible values of these phases:
\begin{equation}
\LL {\cal G} \RR = \prod_{\LA i,j \RA}
\frac{d \varphi_{i,j}}{2 \pi} 
{\cal G}\left( \{ \varphi_{i,j} \} \right)
.
\end{equation}
\item[(ii)] {\sl The phase-resolved
conductance of extended contacts}. This is the
conductance of an extended contact where
the phases $\varphi_{i,j}$ are deterministic
and take the particular
value $\varphi_{i,j} = k_F |{\bf x}_i -  
{\bf x}_j|$ as given by Eq.~\ref{eq:spectral-3D}.
\end{itemize}

Let us consider a model in which a multichannel
fully polarized spin-up electrode is in
contact with a superconductor. At a distance
$D$, there is another fully polarized spin-down
electrode. 
The tight binding model is represented
on Fig.~\ref{fig:deJong}.
There is
one block ``a'' made of $N$ fully polarized spin-up
channels and another block ``b'' made of $M$ fully
polarized spin-down channels. The only difference
with section~\ref{sec:general} is the existence
of a propagator $g_{a_i,a_j}$, $g_{b_i,b_j}$
with $i \ne j$. The form of the
Dyson matrix is still
given by (\ref{eq:Dyson-Matrix}). Compared
to (\ref{eq:X}), there is an additional
summation in the coefficients of the
Dyson matrix:
\begin{eqnarray}
X^{a_i,b_j} &=& \sum_k g^{a_i,a_k}_{1,1}
t^{a_k,\alpha_k} f^{\alpha_k,\beta_j}
t^{\beta_j,b_j}\\
X^{b_i,a_j} &=& \sum_k g^{b_i,b_k}_{2,2}
t^{b_k,\beta_k} f^{\beta_k,\alpha_j}
t^{\alpha_j,a_j}\\
X^{a_i,a_j} &=& \sum_k g^{a_i,a_k}_{1,1}
t^{a_k,\alpha_k} g^{\alpha_k,\alpha_j}
t^{\alpha_j,a_j}\\
X^{b_i,b_j} &=& \sum_k g^{b_i,b_k}_{2,2}
t^{b_k,\beta_k} g^{\beta_k,\beta_j}
t^{\beta_j,b_j}
.
\end{eqnarray}
The derivation of the transport formula
is similar to section~\ref{sec:general}.
For instance,
the subgap Andreev conductance is
the sum of all possible
Cooper pair transmissions:
\begin{equation}
\label{eq:Andreev-multicanal}
{\cal G}^A_{a,b}(\omega) = 4 \pi^2 \sum_{p\in F_a}
\sum_{q \in F_b}
|t^{a_p,\alpha_p}|^2
|t^{b_q,\beta_q}|^2
\rho^{a_p,a_p}_{1,1}(\omega)
\rho^{b_q,b_q}_{2,2}(\omega)
\tilde{f}^{\alpha_p,\beta_q,R}(\omega)
\tilde{f}^{\beta_q,\alpha_p,A}(\omega)
,
\end{equation}
where ``$p \in F_a$'' (``$q \in F_b$'') means that
the $p$ runs over all possible channels
in electrodes a and b.
We used the notation
\begin{eqnarray}
\tilde{f}^{\alpha_p,\beta_q}
&=& \frac{ \tilde{\cal M}^{b_q,a_p}}
{t^{a_p,\alpha_p} t^{b_q,\beta_q}
g^{a_p,a_p}_{1,1} {\cal D}}\\
\tilde{f}^{\beta_q,\alpha_p}
&=& \frac{ \tilde{\cal M}^{a_p,b_q}}
{t^{a_p,\alpha_p} t^{b_q,\beta_q}
g^{b_q,b_q}_{2,2} {\cal D}}
\end{eqnarray}
for the renormalized propagators.
The same 
formalism can be used to handle more
complicated situations involving an arbitrary
number of electrodes having arbitrary
spin polarizations.

%%%%%%%%%%%%%%%%% FIGURE %%%%%%%%%%%%%%%%%%%%%%%%%
\begin{figure}[thb]
\centerline{\fig{8cm}{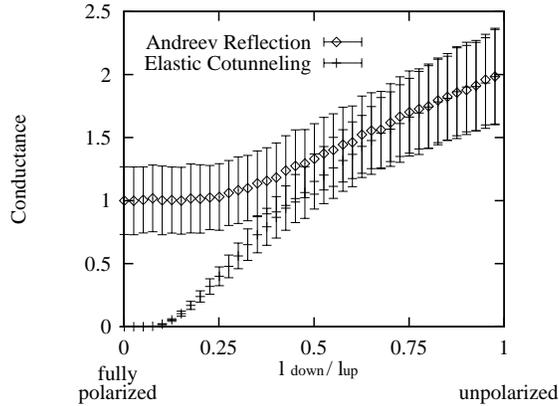}} 
\medskip
\caption{Variation of the $\omega=0$ Andreev and
elastic cotunneling average conductances as a function
of $l^{(\downarrow)}/l^{(\uparrow)}$.
The section of the two electrodes is
circular, with a radius $R=2.1$.
There are $26$ channels in each electrode.
The average conductances are normalized
with respect to the average Andreev conductance
with $\omega=0$ and $l^{(\downarrow)}=0$.
The conductance is distributed as a function
of the phases.
The errorbars indicate the root mean square
of the conductance distribution. 
The parameters are identical to Fig.~\ref{fig:A}.
\base
} 
\label{fig:E}
\end{figure}
%%%%%%%%%%%%%%%%%%%%%%%%%%%%%%%%%%%%%%%%%%%%%%%%%%

\subsection{Andreev reflection {\sl versus}
cotunneling}

Let us consider a system in which two multichannel
electrodes are in contact with a superconductor:
(i) a ferromagnetic electrode; and (ii) a normal metal
electrode with no spin polarization.

Following the discussion in section~\ref{sec:Green-ferro},
we use the ratio $l^{(\downarrow)} 
/ l^{(\uparrow)}$ to parametrize spin polarization.
There is no spin polarization if $l^{(\downarrow)} 
/ l^{(\uparrow)}=1$ and there is a strong
spin polarization if $l^{(\downarrow)} 
/ l^{(\uparrow)} \ll 1$.
We have shown on Fig.~\ref{fig:E} the variation
of the $\omega=0$
Andreev and cotunneling phase averaged conductances
as a function of spin polarization in the
ferromagnetic electrode.
Elastic cotunneling  at $\omega=0$
is vanishingly small if the ferromagnet is
strongly polarized. 
The average Andreev conductance is equal to the
average elastic cotunneling conductance
in the absence
of spin polarization ({\sl i. e.}
with $l^{(\downarrow)} = l^{(\uparrow)}$).

Therefore, the phase averaged conductance
matrix
of the multichannel model behaves like
the phase averaged conductance matrix of the
single channel model.

\subsection{Extended contacts without phase averaging}

\begin{figure}[thb]
\centerline{\fig{8cm}{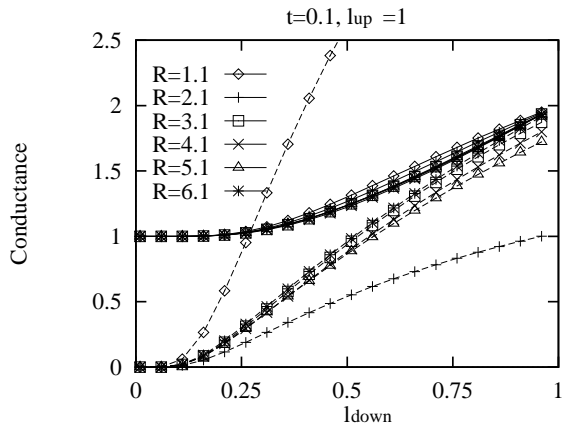}} 
\medskip
\caption{
Variation of the $\omega=0$ Andreev and
elastic cotunneling conductances as a function
of $l^{(\downarrow)}/l^{(\uparrow)}$ for
an extended contact with deterministic phases.
$l^{(\downarrow)}/l^{(\uparrow)}=1$ corresponds
to a non magnetic metal.
The section of the two electrodes is
circular, with a radius $R=1.1$ ($\Diamond$, 10 channels),
$R=2.1$ ($+$, 26 channels), $R=3.1$ ($\Box$, 58 channels),
$R=4.1$ ($\times$, 98 channels), $R=5.1$ ($\Delta$, 178
channels),
$R=6.1$ ($*$, 242 channels). 
The parameters are identical to Fig.~\ref{fig:A}.
The normalizations are identical to Fig.~\ref{fig:E}.
\base
} 
\label{fig:F}
\end{figure}

Now we consider extended contacts in which
the phases take deterministic values.
The phases are given by Eq.~\ref{eq:spectral-3D}:
$\varphi_{i,j} = k_F |{\bf x}_i -  
{\bf x}_j|$. We represented on Fig.~\ref{fig:F}
the dependence of the Andreev and cotunneling
currents as a function of $l^{(\downarrow)} /
l^{(\uparrow)}$ for $\omega=0$.
If the number of channels
is sufficiently large, we see that the
Andreev and elastic cotunneling conductances
are almost identical for the non magnetic metal
($l^{(\downarrow)} /
l^{(\uparrow)}=1$). 
The behavior of extended contacts with deterministic
phases (see Fig.~\ref{fig:F})
is therefore identical to the behavior
of contacts with random phases (see Fig.~\ref{fig:E}).
This result, already established in the
perturbative regime ($\omega \ll \Delta$)
is found to be valid at any frequency and
barrier transparency. We have verified that
this behavior is also valid in the non
perturbative regime ($\omega^* < \omega<\Delta$).
Notice that the same result is found if
only one contact is extended, the other one
having a few channels.

%%%%%%%%%%%%%%%%% FIGURE %%%%%%%%%%%%%%%%%%%%%%%%%
%%%%%%%%%%%%%%%%%%%%%%%%%%%%%%%%%%%%%%%%%%%%%%%%%%

\section{Conclusions}

We have provided in 
this article a detailed theoretical description of ballistic transport in 
multiterminal hybrid structures involving a superconductor and several 
spin-polarized electrodes.  We have performed a non perturbative 
calculation of the conductance matrix, using Keldysh technique, and 
focusing on the spin dependence, the geometry dependence and the energy 
behavior of the two basic processes : Crossed (intercontact) Andreev and 
Elastic Cotunneling.  This generalizes previous perturbative calculations, 
valid only for low contact transparencies and at small voltages.  It also 
generalizes non perturbative calculations made for a single contact and 
using effective Green's functions, which turns out to correspond to a 
certain choice of the phase ($\phi = \pi /2$) in the electronic 
propagators.

A first issue concerns the subgap voltage dependence of the conductances, 
compared to the usual Andreev conductance at a single contact.  A maximum 
is found at a crossover energy (voltage), and the conductance at this 
maximum is reduced compared to the ideal Andreev conductance obtained 
exactly at $\omega = \Delta$ in the single contact case.  The higher the 
transparency of the barriers, the lower is this crossover energy, below 
which the perturbative theory is essentially valid.

The other important issue concerns the phase problem.  Owing to the 
different forms of the normal and anomalous propagators controlling (in the 
superconductor) the cotunneling and Andreev processes respectively, the 
corresponding conductances assume different values for single-channel 
contacts (cotunneling takes place if they have parallel spin polarization, 
crossed Andreev if they have antiparallel ones).  This was exemplified here 
in a three-contact configuration with spin polarizations chosen such as 
both processes can be compared in equivalent geometries.  On the other 
hand, averaging the Fermi phase oscillations present in the propagators 
make the two processes lead to equivalent conductance contributions.  We 
have shown that a self-averaging effect occurs when at least one of the 
contacts has many channels.  The resulting symmetry (equality of 
cotunneling and Andreev crossed conductances) was previously demonstrated 
in a perturbative regime, but it now appears as much more general.  As a 
consequence, if at least one of the contacts is not spin-polarized, the 
resulting intercontact conductance is zero, by compensation of cotunneling 
and Andreev processes~\cite{Falci}.
Conversely, for spin-polarized contacts this offers 
a way of measuring, in amplitude and direction, the polarization of one 
contact with respect to the other through the crossed current measurement, 
as proposed in Ref~\cite{Feinberg-preprint}.  We believe that multiterminal 
superconducting-ferromagnet devices have a large, yet unexplored, potential 
in the growing field of spintronics.  Further theoretical problems concern 
the self-consistent calculation of the superconducting gap~\cite{Apinyan}.

\appendix

\section{Expression of the Keldysh propagator:
two-channel model}
\label{sec:app:keldysh}

In this appendix we derive the expression of the
Keldysh propagator associated to two single-channel
electrodes with $100\%$
spin polarization (see section~\ref{sec:two-channel}).
We need to calculate the Keldysh component:
$\Ht_{a,\alpha}
\HG^{+,-}_{\alpha,a} =
\sum_{i,j} \HK_{i,j}$, with
$i,j \in \{ a,b,\alpha,\beta \}$, and
\begin{eqnarray}
\HK_{\alpha,\alpha} &=&
\Ht_{a,\alpha}
\left[ \HI + \HG^R_{\alpha,a}
\Ht_{a,\alpha} \right]
\Hg_{\alpha,\alpha}^{+,-}
\Ht_{\alpha,a} \HG_{a,a}^A \\
\HK_{\alpha,\beta} &=&
\Ht_{a,\alpha}
\left[ \HI + \HG^R_{\alpha,a}
\Ht_{a,\alpha} \right]
\Hg_{\alpha,\beta}^{+,-}
\Ht_{\beta,b} 
\HG_{b,a}^A\\
\HK_{\beta,\alpha} &=&
\Ht_{a,\alpha}
 \HG_{\alpha,b}^R \Ht_{b,\beta}
\Hg_{\beta,\alpha}^{+,-}
\Ht_{\alpha,a}
\HG_{a,a}^A\\
\HK_{\beta,\beta} &=&
\Ht_{a,\alpha}
\HG_{\alpha,b}^R \Ht_{b,\beta}
\Hg_{\beta,\beta}^{+,-}
\Ht_{b,\beta} \HG^A_{\beta,a}\\
\HK_{a,a}&=&
\Ht_{a,\alpha}
\HG_{\alpha,\alpha}^R \Ht_{\alpha,a}
\Hg_{a,a}^{+,-} \left[ \HI
+ \Ht_{a,\alpha}
\HG_{\alpha,a}^A \right]\\
\HK_{b,b} &=&
\Ht_{a,\alpha}
\HG_{\alpha,\beta}^R \Ht_{\beta,b}
\Hg_{b,b}^{+,-} \Ht_{b,\beta}
\HG_{\beta,b}^A
.
\end{eqnarray}
We also need to calculate
$\Ht_{\alpha,a}
\HG^{+,-}_{a,\alpha} = \sum_{i,j}
\HK'_{i,j}$, with
\begin{eqnarray}
\HK'_{\alpha,\alpha} &=& 
\HG_{a,a}^R \Ht_{a,\alpha}
\Hg^{+,-}_{\alpha,\alpha}
\left[ \HI + \Ht_{\alpha,a}
\HG_{a,\alpha}^A \right]\\
\HK'_{\alpha,\beta} &=&
\HG_{a,a}^R \Ht_{a,\alpha}
\Hg^{+,-}_{\alpha,\beta} 
\Ht_{\beta,b}
\HG_{b,\alpha}^A\\
\HK'_{\beta,\alpha}
&=& \HG_{a,b}^R \Ht_{b,\beta}
\Hg^{+,-}_{\beta,\alpha}
\left[ \HI + \Ht_{\alpha,a}
\HG_{a,\alpha}^A \right]\\
\HK'_{\beta,\beta} &=&
\HG^R_{a,b} \Ht_{b,\beta}
\Hg^{+,-}_{\beta,\beta}
\Ht_{\beta,b} 
\HG^A_{b,\alpha}\\
\HK'_{a,a} &=& \left[ \HI
+ \HG^R_{a,\alpha} \Ht_{\alpha,a}
\right] \Hg^{+,-}_{a,a}
\Ht_{a,\alpha} G^A_{\alpha,\alpha}\\
\HK'_{b,b} &=& \HG_{a,\beta}^R \Ht_{\beta,b}
\Hg^{+,-}_{b,b}
\Ht_{b,\beta} \HG_{\beta,\alpha}^A
.
\end{eqnarray}
The expression of the four terms containing
$n_F(\omega-\mu_S)$ takes the form
\begin{eqnarray}
\label{eq:muS1}
\left[ \HK'_{\alpha,\alpha} -
\HK_{\alpha,\alpha} \right]_{1,1}
&=& 4 \pi^2 n_F(\omega-\mu_S)
|t_{a,\alpha}|^2 \rho^{a,a}_{1,1}
\rho_g^{\alpha,\alpha} \\
\nonb
&& \times \frac{1}{\DA \DR}
\left[ 1 - |t_{b,\beta}|^2
g^{b,b,A}_{2,2}
g^{\beta,\beta,A} \right]
\left[ 1 - |t_{b,\beta}|^2
g^{b,b,R}_{2,2}
g^{\beta,\beta,R} \right]\\
\label{eq:muS2}
\left[ \HK'_{\alpha,\beta} -
\HK_{\alpha,\beta} \right]_{1,1}
&=& 4 \pi^2 n_F(\omega-\mu_S)
|t_{a,\alpha}|^2
|t_{b,\beta}|^2
\rho_{1,1}^{a,a} \rho_f^{\alpha,\beta}
g_{2,2}^{b,b,A}
f^{\beta,\alpha,A}\\
\nonb
&& \times
\frac{1}{\DA \DR} \left[ 1 - 
|t^{b,\beta}|^2 g^{b,b,R}_{2,2}
g^{\beta,\beta,R} \right]\\
\label{eq:muS3}
\left[ \HK'_{\beta,\alpha} -
\HK_{\beta,\alpha} \right]_{1,1}
&=& 4 \pi^2 n_F(\omega-\mu_S)
|t_{a,\alpha}|^2
|t_{b,\beta}|^2
\rho_{1,1}^{a,a} \rho_f^{\beta,\alpha,A}
g_{2,2}^{b,b,R}
f^{\alpha,\beta,R}\\
\nonb
&& \times
\frac{1}{\DA \DR} \left[ 1 - 
|t^{b,\beta}|^2 g^{b,b,A}_{2,2}
g^{\beta,\beta,A} \right]\\
\label{eq:muS4}
\left[ \HK'_{\beta,\beta} -
\HK_{\beta,\beta} \right]_{1,1}
&=& 4 \pi^2 n_F(\omega-\mu_S)
|t_{a,\alpha}|^2
|t_{b,\beta}|^4
\rho_{1,1}^{a,a} \rho_g^{\alpha,\beta}\\
\nonb
&& \times \frac{1}{\DA \DR}
g^{b,b,A}_{2,2} g^{b,b,R}_{2,2}
f^{\alpha,\beta,R} f^{\beta,\alpha,A}
,
\end{eqnarray}
where $\rho_g$ is one of the Nambu components
of the superconductor density of states 
(see Eq.~\ref{eq:Nambu-rho}).
The terms containing $\mu_a$ and $\mu_b$
read
\begin{eqnarray}
\label{eq:mua}
\left[ \HK'_{a,a} -
\HK_{a,a} \right]_{1,1}
&=& -4 \pi^2 n_F(\omega-\mu_a)
|t_{a,\alpha}|^2 \rho^{a,a}_{1,1}
\rho_g^{\alpha,\alpha} \\
\nonb
&& \times \frac{1}{\DA \DR}
\left[ 1 - |t_{b,\beta}|^2
g^{b,b,A}_{2,2}
g^{\beta,\beta,A} \right]
\left[ 1 - |t_{b,\beta}|^2
g^{b,b,R}_{2,2}
g^{\beta,\beta,R} \right]\\
\nonb
%%%%%%%%%%%%%%%%%%%%%%%%%%%%%%%%%%%%
&+& 2 i \pi n_F(\omega-\mu_a)
|t_{a,\alpha}|^2 |t_{b,\beta}|^2
\rho^{a,a}_{1,1}\\
\nonb
&& \times \frac{1}{\DA \DR}
g^{b,b,A}_{2,2} 
f^{\alpha,\beta,A} f^{\beta,\alpha,A}
\left[ 1 - |t_{b,\beta}|^2
g^{b,b,R}_{2,2}
g^{\beta,\beta,R} \right]\\
%%%%%%%%%%%%%%%%%%%%%%%%%%%%%%%%%%%%
\nonb
&-& 2 i \pi n_F(\omega-\mu_a)
|t_{a,\alpha}|^2 |t_{b,\beta}|^2
\rho^{a,a}_{1,1}\\
\nonb
&& \times \frac{1}{\DA \DR} g^{b,b,R}_{2,2} 
f^{\alpha,\beta,R} f^{\beta,\alpha,R}
\left[ 1 - |t_{b,\beta}|^2
g^{b,b,A}_{2,2}
g^{\beta,\beta,A} \right] \\
%%%%%%%%%%%%%%%%%%%%%%%%%%%%%%%%%%%%
\label{eq:mub}
\left[ \HK'_{b,b} -
\HK_{b,b} \right]_{1,1}
&=& 4 \pi^2 n_F(\omega-\mu_b)
|t_{a,\alpha}|^2
|t_{a',\alpha'}|^2
\rho^{a,a}_{1,1}
\rho^{b,b}_{2,2}
\frac{1}{\DA \DR}
f^{\alpha,\beta,R} f^{\beta,\alpha,A}
.
\end{eqnarray}
One arrives at the identity
\begin{eqnarray}
&& \frac{1}{n_F(\omega-\mu_S)}
\left[ \HK_{\alpha,\alpha}' - \HK_{\alpha,\alpha}
+\HK_{\alpha,\beta}' - \HK_{\alpha,\beta}
+\HK_{\beta,\alpha}' - \HK_{\beta,\alpha}
+\HK_{\beta,\beta}' - \HK_{\beta,\beta}\right]_{1,1}\\
&=&
-\frac{1}{n_F(\omega-\mu_a)}
\left[ \HK_{a,a}' - \HK_{a,a} \right]_{1,1}
- \frac{1}{n_F(\omega-\mu_b)}
\left[ \HK_{b,b}' - \HK_{b,b} \right]_{1,1}
,
\end{eqnarray}
which constitutes for this particular system
a proof of the trick on the spectral
current (see section~\ref{sec:conj-spectral}).
The expression of the Keldysh propagators given
in this appendix can be used to obtain the
transport formula given in section~\ref{sec:4.2}.

\section{Derivation of the transport formula
with an arbitrary number of single-channel electrodes}
\label{app:B}

We present in this appendix the derivation of the
transport formula given by
Eqs.~(\ref{eq:transport-gene})~--~(\ref{eq:tr-gen-AR}),
associated to a situation where $N$ ferromagnetic
spin-up electrodes and $M$ ferromagnetic
spin-down electrodes are in contact
with a superconductor (see Fig.~\ref{fig:schema3}).

\subsection{Solution of the Dyson equation}
The unknown Green's functions
$\{ G^{a_1,a_1}, ..., G^{a_N,a_1},
G^{b_1,a_1}, ..., G^{b_M,a_1} \}$ are
the solution of the Dyson equation
$$
\hat{\cal M} \left[
\begin{array}{c}
G^{a_1,a_1} \\
G^{a_2,a_1} \\
... \\
G^{a_N,a_1} \\
G^{b_1,a_1} \\
... \\
G^{b_M,a_1}
\end{array} \right]
= \left[
\begin{array}{c}
g^{a_1,a_1} \\
0 \\
... \\
0 \\
0 \\
... \\
0 \end{array} \right]
,
$$
where the Dyson matrix $\hat{\cal M}$ takes the form
\begin{equation}
\label{eq:Dyson-Matrix}
\hat{\cal M} = \HI +
\left[
\begin{array}{cc}
- \HY^{a,a} & \HX^{a,b} \\
\HX^{b,a} & - \HY^{b,b}
\end{array}
\right]
,
\end{equation}
where $\HY^{a,a}$ is a $N \times N$
block, $\HX^{a,b}$ is a $N \times M$
block
The matrix elements of $\HX$ and $\HY$
are 
\begin{eqnarray}
\label{eq:X}
X^{a_i,a_j} &=& t^{a_i,\alpha_i}
t^{a_j,\alpha_j} g^{a_i,a_j}
f^{\alpha_i,\alpha_j} \\ 
Y^{a_i,a_j} &=& t^{a_i,\alpha_i}
t^{a_j,\alpha_j} g^{a_i,a_i} 
g^{\alpha_i,\alpha_j}
\label{eq:Y}
.
\end{eqnarray}
The solution of the Dyson equation takes the
form
\begin{eqnarray}
G^{a_i,a_j}_{1,1} &=&
\frac{(-)^{i+j}}{\cal D} g^{a_j,a_j} \tilde{\cal M}_{a_j,a_i} \\
G^{b_i,b_j}_{2,2} &=&
\frac{(-)^{i+j}}{\cal D} g^{b_j,b_j} \tilde{\cal M}_{b_j,b_i} \\
G^{a_i,b_j}_{1,2} &=&
\frac{(-)^{i+j+N}}{\cal D} g^{b_j,b_j} \tilde{\cal M}_{b_j,a_i} \\
G^{b_i,a_j}_{2,1} &=&
\frac{(-)^{i+j+N}}{\cal D} g^{a_j,a_j} \tilde{\cal M}_{a_j,b_i} 
,
\end{eqnarray}
where ${\cal D}$ is the determinant of the Dyson
matrix and $\tilde{\cal M}_{a_i,a_j}$
are the minors of this matrix.

%%%%%%%%%%%%%%%%%%%%%%%%%%%%%%%%%%%%%%%%%%%%%%%%%%%%%%%%%%%%%
\subsection{Solution of the Dyson-Keldysh equation}
To obtain the current through the link
$a_1$~--~$\alpha_1$, we need to evaluate
the Keldysh component
\begin{eqnarray}
\label{eq:Gpm-gen-1}
\Ht^{a_1,\alpha_1} G^{+,-}_{\alpha_1,a_1} &=& 
\Ht^{a_1,\alpha_1}
\HG^{\alpha_1,\alpha_1,R}
\Ht^{\alpha_1,a_1}
\Hg^{+,-}_{a_1,a_1}
\left[ \HI + \Ht^{a_1,\alpha_1}
\HG^{\alpha_1,a_1,A} \right] \\
\label{eq:Gpm-gen-2}
&+& \sum_{k=2}^{N} 
\Ht^{a_1,\alpha_1}
\HG^{\alpha_1,\alpha_k,R}
\Ht^{\alpha_k,a_k}
\Hg^{+,-}_{a_k,a_k}
\Ht^{a_k,\alpha_k}
\HG^{\alpha_k,a_1,A}\\
\label{eq:Gpm-gen-3}
&+& \sum_{k=1}^N
\Ht^{a_1,\alpha_1}
\HG^{\alpha_1,\beta_k,R}
\Ht^{\beta_k,b_k}
\Hg^{+,-}_{b_k,b_k}
\Ht^{b_k,\beta_k}
\HG^{\beta_k,a_1,A}
.
\end{eqnarray}
Let us start with (\ref{eq:Gpm-gen-1}).
The first step is to show that
$$
(\ref{eq:Gpm-gen-1}) = 2 i \pi
n_F(\omega-\mu_{a_1}) |t^{a_1,\alpha_1}|^2
\rho_{1,1}^{a_1,a_1}
G_{1,1}^{\alpha_1,\alpha_1,R}
\left[ \HI + \Ht^{a_1,\alpha_1}
\HG^{\alpha_1,\alpha_1,R}\right]_{1,1}
.
$$
The different terms in this equation
are found to be
$$
\left[ \HI + \Ht^{a_1,\alpha_1}
\HG^{\alpha_1,\alpha_1,R}\right]_{1,1}
= \frac{ \tilde{\cal M}_{a_1,a_1}}{\cal D}
,
$$
and
\begin{eqnarray}
|t^{a_1,\alpha_1}|^2 G^{\alpha_1,\alpha_1}
&=& |t^{a_1,\alpha_1}|^2 \frac{1}{\cal D}
g^{\alpha_1,\alpha_1} \tilde{\cal M}_{a_1,a_1}\\
&+& \frac{1}{\cal D} \sum_{k \ne 1}
(-)^{k+1} t^{a_1,\alpha_1} t^{a_k,\alpha_k}
g^{\alpha_1,\alpha_k}
\tilde{\cal M}_{a_1,a_k}\\
&+& \frac{1}{\cal D} \sum_{k=1}^M
(-)^{k+N} t^{a_1,\alpha_1}
t^{b_k,\beta_k}
f^{\alpha_1,\beta_k}
\tilde{\cal M}_{a_1,b_k}
.
\end{eqnarray}
To evaluate (\ref{eq:Gpm-gen-2}),
we first show that 
$$
(\ref{eq:Gpm-gen-2}) = \sum_{k=2}^N
2 i \pi n_F(\omega-\mu_{a_k})
t^{a_1,\alpha_1} t^{a_k,\alpha_k}
\rho_{1,1}^{a_k,a_k}
G_{1,1}^{\alpha_1,\alpha_k,R}
\left[ \Ht^{a_k,\alpha_k}
G^{\alpha_k,a_1} \right]_{1,1}^A
.
$$
Using the identities
\begin{eqnarray}
\label{eq:iden-gen1}
\left[ \Ht^{a_k,\alpha_k}
G^{\alpha_k,a_1} \right]_{1,1} &=&
(-)^{k+1} \frac{ g^{a_1,a_1}}{g^{a_k,a_k} {\cal D}}
\tilde{\cal M}_{a_1,a_k}\\
\label{eq:iden-gen2}
\left[ G^{\alpha_1,\alpha_k} \right]_{1,1}
&=& \frac{(-)^{k+1}}{t^{a_1,\alpha_1} t^{a_k,\alpha_k}
g^{a_1,a_1} {\cal D} } \tilde{\cal M}_{a_k,a_1}
,
\end{eqnarray}
we obtain
$$
(\ref{eq:Gpm-gen-2}) = \sum_{k=2}^N
2 i \pi n_F(\omega-\mu_{a_k})
t^{a_1,\alpha_1}
t^{a_k,\alpha_k}
\rho^{a_k,a_k}_{1,1} g^{a_1,a_1,A}
\tilde{g}_{\alpha_1,\alpha_k}^R
\tilde{g}_{\alpha_k,\alpha_1}^A
,
$$
where $\tilde{g}_{\alpha_1,\alpha_k}$
denotes a renormalized propagator.
We use a similar calculation
to evaluate (\ref{eq:Gpm-gen-3}) and
we deduce the transport formula given by
Eqs.~(\ref{eq:transport-gene})~--~(\ref{eq:tr-gen-AR}).


\begin{thebibliography}{99}
\bibitem{Martin} G.B. Lesovik, T. Martin, and G.
Blatter, arXiv:cond-mat/0009193.

\bibitem{Loss} M. S. Choi, C. Bruder and D. Loss,
Phys. Rev. B {\bf62}, 13569 (2000);

P. Recher, E. V. Sukhorukov and D. Loss,
Phys. Rev. B {\bf63}, 165314 (2001).

\bibitem{Feinberg} G. Deutscher and D. Feinberg,
App. Phys. Lett. {\bf 76},
487 (2000).

\bibitem{Falci} G. Falci, D. Feinberg, and F.W.J. Hekking,
Europhysics Letters {\bf 54}, 255 (2001).

\bibitem{Melin} R. M\'elin, J. Phys.: Condens. Matter
{\bf 13}, 6445 (2001);

R. M\'elin, to appear in the Proceedings of the
XXXVIth Rencontres de Moriond, T. Martin and 
G. Montambaux Eds., EDP Sciences (2001).

\bibitem{deJong} M.J.M. de Jong and
C.W.J. Beenakker, Phys. Rev. Lett.
{\bf 74}, 1657 (1995).

\bibitem{Soulen} R.J.~Soulen {\sl et al.},
Science {\bf 282}, 85 (1998).

\bibitem{Upadhyay} S.K.~Upadhyay {\sl et al.},
Phys. Rev. Lett. {\bf 81}, 3247 (1998).

\bibitem{Buzdin} I. Baladi\'e, A. Buzdin,
N. Ryzhanova, and A. Vedyayev,
Phys. Rev. B {\bf 63}, 054518 (2001).

A. Buzdin, A.V. Vedyayev, and N. Ryzhanova,
Europhys. Lett. {\bf 48}, 686 (1999).

\bibitem{Apinyan} V. Apinyan and
R. M\'elin, cond-mat/0107038.

\bibitem{Melin00} R. M\'elin, Europhysics
Letters {\bf 51}, 202 (2000).

\bibitem{Giroud} M. Giroud {\sl et al.},
Phys. Rev. B {\bf 58}, R11872 (1998).

\bibitem{Petrashov} V.T. Petrashov, I.A. Sosnin,
and C. Troadec, arXiv:cond-mat/0007278.

\bibitem{Chandra} J. Aumentado and V. Chandrasekhar,
arXiv:cond-mat/0007433.

\bibitem{Bauer} W. Belzig, A. Brataas, Yu. V. Nazarov,
and G.E. Bauer, arXiv:cond-mat/0005188.

\bibitem{Byers} J.M. Byers and M.E. Flatt\'e,
Phys. Rev. Lett. {\bf 74}, 306 (1995).

\bibitem{Cuevas}
J.C. Cuevas, A. Martin-Rodero, and A.
Levy Yeyati, Phys. Rev. B {\bf 54}, 7366 (1996).

\bibitem{Keldysh} L.V. Keldysh, Sov. Phys.
JETP {\bf 20}, 1018 (1965).

\bibitem{Caroli}  C. Caroli, R. Combescot,
P. Nozi\`eres and D. Saint-James,
J. Phys. C: Solid St. Phys. {\bf 4}, 916 (1971);
{\sl ibid.} {\bf 5}, 21 (1972).

\bibitem{Valet-Fert} 
T. Valet and A. Fert, Phys. Rev. B {\bf 48}, 7099 (1993);

T. Valet and A. Fert, J. Magn. Magn. Mater. {\bf 121},
378 (1993);

A. Fert, T. Valet, and J. Barnas,
J. Appl. Phys. {\bf 75}, 6693 (1994);

A. Fert, J.L. Duvail, and T. Valet,
Phys. Rev. B {\bf 52}, 6513 (1995).

\bibitem{Gijs-Bauer} M.A.M. Gijs and G. Bauer,
Adv. in Physics {\bf 46}, 285 (1997).

\bibitem{Melin-Denaro} R. M\'elin and D. Denaro,
Eur. Phys. J. B {\bf 18}, 149 (2000).

\bibitem{BTK} G.E. Blonder, M. Tinkham, and
T.M. Klapwijk, Phys. Rev. B {\bf 25},
4515 (1982).

\bibitem{Feinberg-preprint} D. Feinberg,
G. Deutscher, G. Falci and F. Hekking,
to appear in the Proceedings of the
XXXVIth Rencontres de Moriond, T. Martin and 
G. Montambaux Eds., EDP Sciences (2001).



\end{thebibliography}
\end{document}